\documentclass[pra,amsmath,amssymb,aps,10pt,groupedaddress,superscriptaddress,longbibliography,twocolumn,floatfix]{revtex4-2}

\usepackage{amsmath, amsthm, amssymb, amsfonts}
\usepackage{graphicx} 
\usepackage[colorlinks=true, linkcolor=blue, urlcolor=blue, citecolor=blue, anchorcolor = blue]{hyperref} 
\usepackage{color,xcolor}
\usepackage{orcidlink}
\usepackage[utf8]{inputenc}
\usepackage[T1]{fontenc} 

\usepackage{dcolumn}
\newcolumntype{d}[1]{D{.}{.}{#1}}

\usepackage{soul} 
\usepackage[normalem]{ulem}

\definecolor{orange}{rgb}{0.99,0.5,0.31}

\newlength{\lengthofminus}
\newcommand{\dl}{\hspace{\lengthofminus}} 

\newlength{\lengthofminusscriptsize}
\newcommand{\dls}{\hspace{\lengthofminusscriptsize}} 

\newlength{\lengthofone}
\newcommand{\dd}{\hspace{\lengthofone}} 

\begin{document}
\title{Precision measurement of Cs($nF_{J}$) quantum defects and calculations of scalar and tensor polarizabilities of the $nS_{1/2}$, $nP_J$ ,$nD_J$, and $nF_J$ series}


\author{Pinrui Shen}
\affiliation{ Quantum Valley Ideas Laboratories, 485 Wes Graham Way, Waterloo, Ontario \\N2L 6R1, Canada
}
\author{Mariusz Pawlak\orcidlink{0000-0002-2200-8287}}
\affiliation{ 
Faculty of Chemistry, Nicolaus Copernicus University in Toru\'n, Gagarina~7, 87-100~Toru\'n, Poland
}
\author{Donald Booth}
\affiliation{ Quantum Valley Ideas Laboratories, 485 Wes Graham Way, Waterloo, Ontario \\N2L 6R1, Canada
}
\author{Kent Nickerson}
\affiliation{ Quantum Valley Ideas Laboratories, 485 Wes Graham Way, Waterloo, Ontario \\N2L 6R1, Canada
}

\author{Haddad Miladi}
\affiliation{ Quantum Valley Ideas Laboratories, 485 Wes Graham Way, Waterloo, Ontario \\N2L 6R1, Canada
}
\author{H. R. Sadeghpour\orcidlink{0000-0001-5707-8675}}
\email{e-mail: hsadeghpour@cfa.harvard.edu}
\affiliation{ITAMP, Center for Astrophysics $|$ Harvard \& Smithsonian, 60 Garden St., Cambridge, Massachusetts 02138, USA
}
\author{James Shaffer}
\email{e-mail: jshaffer@qvil.ca}
\affiliation{ Quantum Valley Ideas Laboratories, 485 Wes Graham Way, Waterloo, Ontario \\N2L 6R1, Canada
}

\begin{abstract}
In this paper, we extend our recent work on cesium $S$ and $D$ states [Phys. Rev. Lett. 133, 233005 (2024)] to the $F$ states. We present absolute frequency measurements of the $|6S_{1/2}, F=3\rangle \rightarrow nF_{5/2,7/2}(n=28$--$68)$ Rydberg series to measure the spectrum of $^{133}$Cs. Atomic spectra are obtained using a three-photon excitation scheme referenced to an optical frequency comb in a sample of ultracold $^{133}$Cs. By globally fitting the absolute-frequency measurements to the  modified Ritz formula, we determine the quantum defects of the $nF_{5/2}$ and $nF_{7/2}$ series. The ionization potential extracted for both series from the modified Ritz formula agrees with our measurements based on the $S$ and $D$ series. Fine-structure intervals are calculated and parameterized. The wave functions computed for the energies from the quantum defects are used to calculate transition dipole moments. We compare the reduced electric-dipole matrix elements with available benchmarks and find agreement within the precision of those works. The scalar and tensor polarizabilities of the $nS_{1/2}$, $nP_J$, $nD_J$ and $nF_J$ series are calculated based on the now more accurate set of wave functions. Moreover, we report the polarizability as a series in powers of the effective principal quantum number and find the main coefficients of the expansion. The results will be useful for calculating properties of $^{133}$Cs such as collision and decay rates, polarizabilities, and magic wavelengths.
\end{abstract}

\maketitle

\section{Introduction}

Precise measurements of atomic energy levels are necessary for determination of a number of fundamental constants in physics~\cite{McIntyre1988, Safronova2018, Scheidegger2024} in addition to providing stringent tests for the accuracy of atomic structure calculations, including quantum defect theories. Rapid developments in quantum measurement technologies, such as optical clocks~\cite{Martin2018,Ludlow2015,Pan2020} and Rydberg electric field sensors~\cite{Fan2015,Nowosielski2024,Sedlacek2013}, are adding impetus for the accurate determination of the energy levels and electric field polarizabilities in cold and hot vapor-atomic systems.

In our recent work on the energy levels of cesium~\cite{Shen2024}, we precisely measured the energy levels of $nS_{1/2}$ and $nD_J$ Rydberg states and reported updated quantum defects as well as the ionization energy for cesium. We evaluated the core contributions for the $nS_{1/2}$ and $nD_J$ series and demonstrated the significant role of core penetration. One of the goals of this work is to confirm the reduction of the core  penetration effect for higher angular momentum states, here the $nF_J$ states. For $nF_J$, the core polarization dominates the core contributions.

We used a three-photon excitation scheme to perform precision measurements of the energy levels of $^{133}$Cs($nF_{5/2}$) and $^{133}$Cs($nF_{7/2}$) for $n=28$--68 with an accuracy of $<60$~kHz. We determine the quantum defects and the ionization energy of the $nF_{5/2}$ and $nF_{7/2}$ series by fitting the absolute-frequency measurements to the modified Ritz formula. The ionization energies extracted from the $nF_J$ series agree with our previous work~\cite{Shen2024}. We compile new measurements of energy levels and quantum defect parameters, which, when combined with the measurements of the Cs($nP_J$) states~\cite{Deiglmayr2016}, encompass the $nS_{1/2}$, $nP_J$, $nD_J$ and $nF_J$ Rydberg series of non-hydrogenic states in Cs.

We numerically determine the wave functions based on the one-electron model potential and the energies from the quantum defects. Next, we calculate the reduced dipole matrix elements of the transitions $nD_J\leftrightarrow n'F_{J'}$ and the scalar and tensor polarizabilities of the $nS_{1/2}$, $nP_J$, $nD_J$, and $nF_J$ series. The results presented in this work represent the most accurate compilation of the energy levels, radiative matrix elements, and polarizabilities of a~wide range of Rydberg excitations in cesium. They are compared with recent experimental and theoretical data, showing the high quality of our findings. Importantly, they surpass the results from the ARC (Alkali Rydberg Calculator) library~\cite{ARC2017,ARCv3_2020} in many instances.


\section{Experimental Setup}\label{EXPapparatus}

\begin{figure}[!b]
    \centering
    \includegraphics[width=0.48\textwidth]{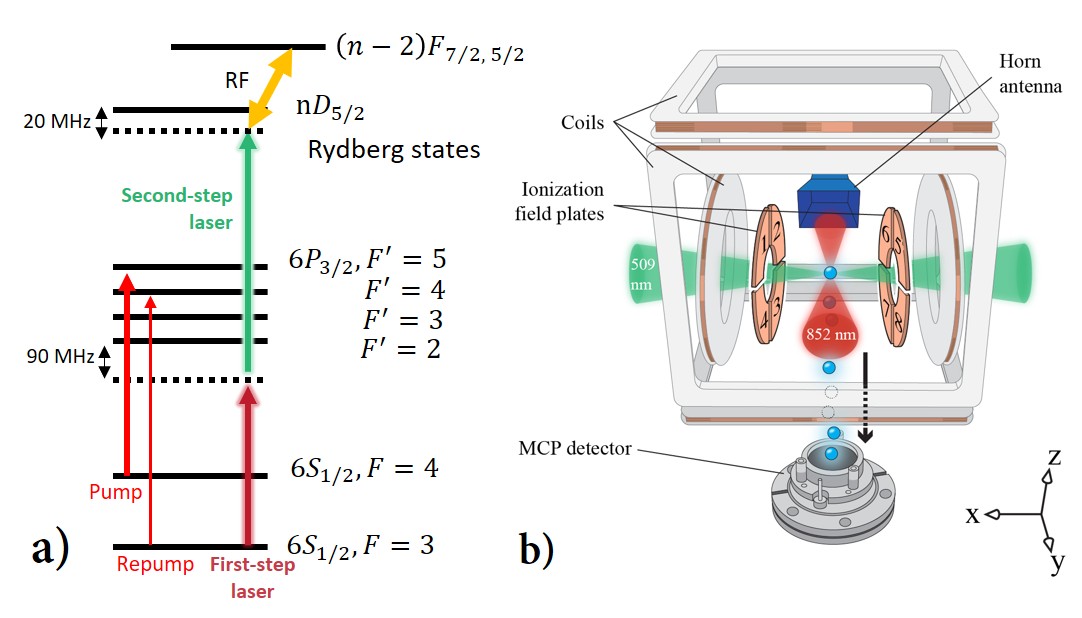}
    \caption{Schematic of the experiment: (a) The transitions driven by the MOT trapping lasers, the excitation lasers, and the RF radiation; (b) The key components of the apparatus, including the ring-shaped electrodes, the horn antenna, the flight tube, and the MCP detector. The excitation lasers are aligned perpendicular to each other in order to minimize the excitation volume. The RF field is transmitted to the atom cloud via a~horn antenna. The frequency of the RF field is varied from 9 to 133~GHz to couple different $nD_{5/2}\rightarrow n'F_J$ transitions.}
    \label{fig:expill}
\end{figure}

A magneto-optical trap (MOT) is used to cool and trap a sample of atoms. The MOT cooling laser is frequency locked 10 MHz below the $D_2$ ($|6S_{1/2}, F=4\rangle \rightarrow |6P_{3/2},F'=5\rangle$) transition.  The MOT repump laser is sideband-generated and in resonance with the $|6S_{1/2}, F=3\rangle \rightarrow |6P_{3/2},F'=4\rangle$ transition. Figure~\ref{fig:expill}(a) shows the MOT lasers and their frequencies. The axial magnetic field gradient for the MOT is  $12~\mathrm{G}\,\mathrm{cm}^{-1}$. Two segmented, ring-shaped electrodes spaced by $4~\rm{cm}$ surrounding the cold Cs cloud are used to cancel background DC electric fields as well as ionize Rydberg atoms, as shown in Fig.~\ref{fig:expill}(b). Each ring-shaped electrode is made of four separate copper field plates that can be supplied with independent voltages. A~high voltage pulse is applied to the top four plates to ionize the Rydberg atoms and project them onto a microchannel plate (MCP) detector. Outside the MOT vacuum chamber, three pairs of Helmholtz coils are placed to compensate stray magnetic fields.

The measurement cycle starts by cooling and trapping Cs atoms in the MOT. After loading Cs atoms in the MOT for 5 seconds, a polarization gradient cooling (PGC) step is performed to further reduce the temperature of the atoms to $\sim 2\,\mu$K~\cite{Shen2024}. After the PGC step, the repump beam is extinguished and the cooling laser is applied for an additional 2$\,$ms so that the Cs atoms are optically pumped to the $|6S_{1/2}, F=3\rangle$ ground hyperfine state.

Two excitation lasers and a radio frequency (RF) photon are used to excite Cs atoms from the $|6S_{1/2}, F=3\rangle$ state to $nF_{J}$ Rydberg states. A first-excitation step laser at 852~nm and a second-excitation step laser at 509~nm are both locked to a frequency comb that is locked to a GPS steered Rb clock, resulting in laser linewidths of less than 200~Hz and an absolute frequency stability of $2\times10^{-12}/\sqrt{\rm{Hz}}$. The RF field is generated by an RF oscillator, which is also locked to the GPS steered Rb clock. The RF field is transmitted to the atom cloud via a horn antenna. The first-step laser is red-detuned 90~MHz below the transition from the $|6S_{1/2}, F=3\rangle$ ground hyperfine state to the $|6P_{3/2}$, $F'=2\rangle$ excited state. The second-step laser is 70~MHz blue-detuned from the $|6P_{3/2}$, $F'=2\rangle$ to the $nD_{5/2}$ Rydberg state to adiabatically eliminate the $nD_{5/2}$ states. The RF field is 20~MHz blue-detuned from the $nD_{5/2}$ to $(n-2)F_{5/2,7/2}$ Rydberg transition. The two excitation lasers are crossed in a perpendicular geometry, both with parallel linear, vertical polarizations. The Gaussian beam waists for the first-step laser and the second-step laser are $\omega_0=3~\rm{mm}$ and $\omega_0=400~\mu\rm{m}$, respectively.  The duration of the excitation laser pulses and RF pulses are 30$\,\mu\rm s$. The pulses overlap in time.

To perform the Rydberg state spectroscopy measurements, the frequencies of the first- and second-step lasers are fixed and the frequency of the RF field is scanned. Following the excitation pulses, a 2$\,\mu\rm s$ high voltage pulse is applied to ionize the Rydberg atoms and project them onto a multichannel plate (MCP) detector. Each ion pulse is recorded by a multichannel analyzer. We adjust the power of excitation fields to achieve the average number of excited Rydberg atoms per shot below two atoms in order to minimize the effect from ions and Rydberg atom interactions. The Rydberg excitation and ion detection sequence is repeated 40 times for each cold atomic cloud. Twenty clouds are used for each RF frequency, resulting in 800 excitation-detection cycles per RF frequency step.

\section{Results and uncertainty analysis}

\begin{figure}[!ht]
    \centering
    \includegraphics[width=0.48\textwidth]{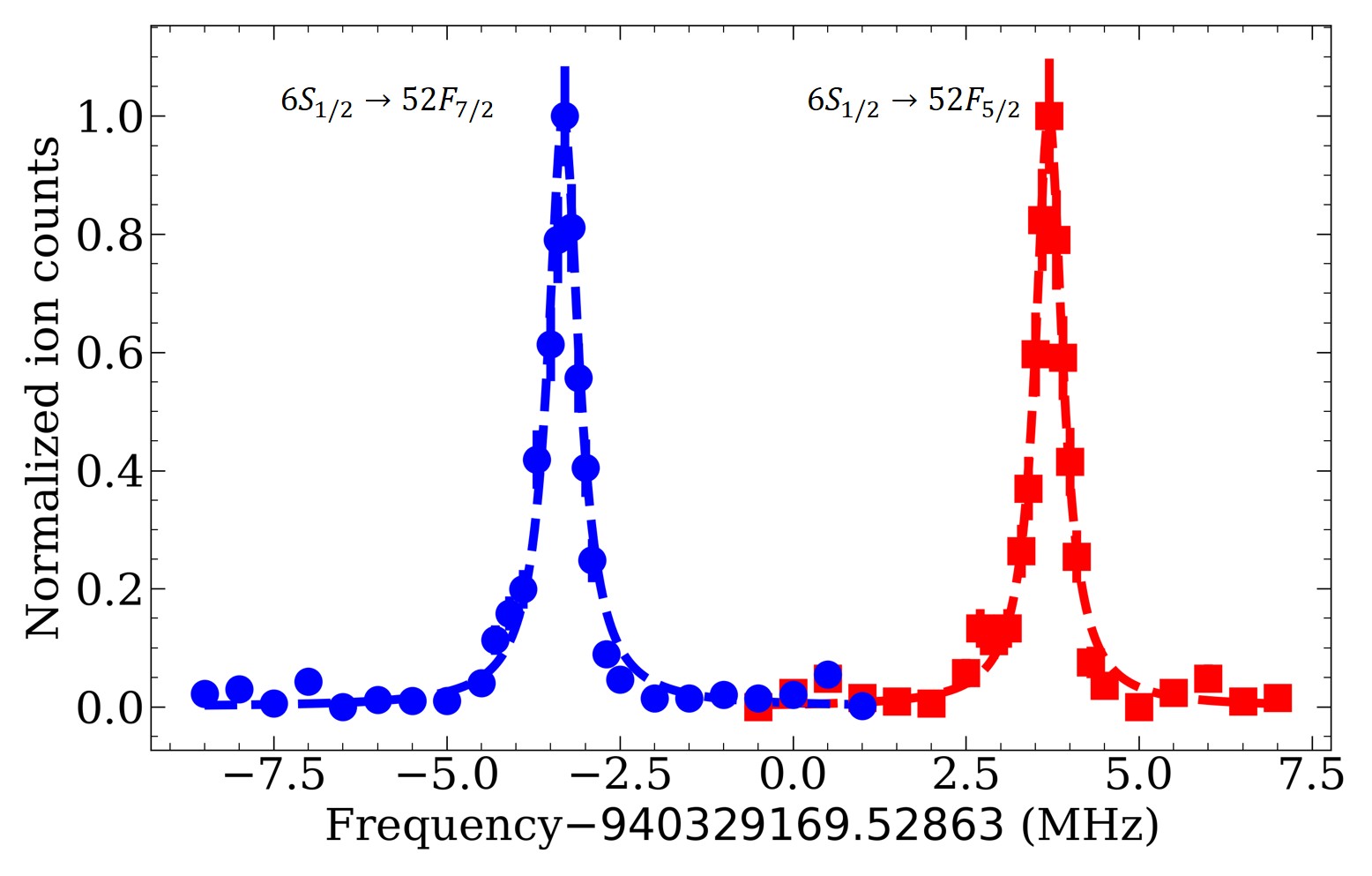}
    \caption{An example of a measured spectrum of the $|6S_{1/2}, F=3\rangle \rightarrow 52F_{5/2,7/2}$ transition. The spectrum is fit to a single Lorentzian function and yields a linewidth of $\sim200~\mathrm{kHz}$. The $52F_{7/2}$ energy level is lower than the $52F_{5/2}$ energy level due to the inversion of the fine structure of the Cs($nF_{J}$) levels. The vertical axis represents the number of Rydberg ions normalized to its peak value. The error bars represent the shot-to-shot uncertainty of each data point. The horizontal axis indicates the transition frequency from the $|6S_{1/2}, F=3\rangle$ state to the target Rydberg state, $52F_{5/2,7/2}$. }
    \label{fig:expill2}
\end{figure}

Absolute frequency measurements are performed for both the $nF_{5/2}$ and $nF_{7/2}$ series ($n=28$--$68$). An example of the measured spectrum of the $52F_{5/2,7/2}$ Rydberg state energy is shown in Fig.~\ref{fig:expill2}. Due to the fine-structure inversion of the cesium $nF_J$ states, the $52F_{7/2}$ energy level is lower that of the $52F_{5/2}$ level. Because the hyperfine splittings of the $nF_{J}$ states are smaller than the statistical uncertainty, a single Lorentzian fit is used to fit the spectrum. The center frequency extracted from the fit equals the sum of the frequencies of the first-step laser, the second-step laser and the RF field and represents a weighted average over unresolved hyperfine states. To account for the hyperfine splitting effect, we reference the measured transition frequencies to the center of gravity of the Rydberg level and provide the uncertainty for the hyperfine splitting. The absolute frequency measurements for the $nF_{5/2,7/2} (n=28$--$68)$ Rydberg  series are presented in Table~\ref{tab:absf_measurements}.

\begin{table}[!t]
\centering
\caption{Absolute frequency measurements and the statistical uncertainties (in brackets) of transitions from the $|6S_{1/2}, F=3 \rangle$ hyperfine ground state to the center of gravity of the $nF_{7/2}$ and $nF_{5/2}$ states, accounting for the uncertainty due to the unresolved hyperfine splittings, see the main text.
}
\begin{small}    
    \begin{tabular}{c|l|l}
    \hline
    \hline
        $n$ & \multicolumn{1}{c|}{$\nu_{F_{7/2}}$ (kHz)} & \multicolumn{1}{c}{$\nu_{F_{5/2}}$ (kHz)} \\ \hline
         28 & 937336006858.3(13.4)    & 937336050416.9(16.2) \\
         29 & 937621405933.9(7.6)     & 937621445185.3(8.7)  \\
         30 & 937878713869.0(13.5)    & 937878749474.3(11.1) \\
         31 & 938111499897.7(17.8)    & 938111532039.6(15.2) \\
         32 & 938322783490.2(10.1)    & 938322812859.1(7.3)  \\
         33 & 938515133646.6(19.8)    & 938515160290.5(14.3) \\
         34 & 938690746612.2(19.3)    & 938690771062.7(18.4) \\
         35 & 938851509582.6(12.8)    & 938851532073.1(16.1) \\
         36 & 938999050969.2(16.6)    & 938999071413.1(10.1) \\
         37 & 939134781504.5(8.6)     & 939134800504.8(19.5) \\
         38 & 939259928990.1(13.2)    & 939259946508.2(7.2)  \\
         39 & 939375565925.6(22.7)    & 939375581984.5(17.1) \\
         40 & 939482631848.9(11.7)    & 939482647095.1(7.9)  \\
         41 & 939581953717.1(16.8)    & 939581967693.6(9.9)  \\
         42 & 939674260327.3(13.8)    & 939674273416.2(12.1) \\
         43 & 939760197476.4(21.5)    & 939760209691.2(10.5) \\
         44 & 939840337734.5(14.1)    & 939840349041.3(17.1) \\
         45 & 939915190973.8(19.4)    & 939915201670.7(13.1) \\
         46 & 939985212321.4(9.7)     & 939985222355.4(11.3) \\
         47 & 940050808852.3(7.5)     & 940050818387.9(11.8) \\
         48 & 940112345863.2(14.1)    & 940112354580.7(10.4) \\
         49 & 940170151278.9(5.7)     & 940170159551.9(7.0)  \\
         50 & 940224521028.7(15.4)    & 940224528606.5(7.6)  \\
         51 & 940275721760.3(9.6)     & 940275729082.6(13.8) \\
         52 & 940323995446.8(10.1)    & 940324002408.9(8.9)  \\
         53 & 940369560752.2(21.5)    & 940369567207.8(13.3) \\
         54 & 940412616583.6(13.1)    & 940412622797.2(9.1)  \\
         55 & 940453344017.7(8.3)     & 940453349894.7(15.9) \\
         56 & 940491907850.0(9.1)     & 940491913334.4(7.2)  \\
         57 & 940528458753.1(12.1)    & 940528463937.7(15.5) \\
         58 & 940563134272.9(7.5)     & 940563139277.9(9.5)  \\
         59 & 940596060712.9(8.8)     & 940596065434.1(10.4) \\
         60 & 940627353554.7(7.6)     & 940627358173.1(14.9) \\
         61 & 940657119447.6(7.9)     & 940657123795.0(10.8) \\
         62 & 940685455781.1(6.1)     & 940685459986.7(13.5) \\
         63 & 940712452995.8(10.6)    & 940712456750.4(10.9) \\
         64 & 940738193903.3(10.3)    & 940738197442.9(8.7)  \\
         65 & 940762755231.6(9.2)     & 940762758768.1(17.5) \\
         66 & 940786208097.4(11.4)    & 940786211475.1(17.9) \\
         67 & 940808618294.9(15.9)    & 940808621446.2(10.8) \\
         68 & 940830046480.2(14.6)    & 940830049498.7(11.9) \\
         \hline \hline  
    \end{tabular}
    \label{tab:absf_measurements}
\end{small}
\end{table}

In Table~\ref{tab:absf_measurements}, the bracket in each column represents the statistical uncertainty. The dominant source of statistical uncertainty is the shot-to-shot variations in the Rydberg ion counts. The uncertainty in the number of ions is $< 15 \%$ and contributes around $10~\rm{kHz}$ to the total statistical uncertainty in the absolute frequency determination. A~second contribution to the statistical uncertainty of the line position is the uncertainty in the laser frequencies. The two excitation lasers are both locked to a~frequency comb which has a stability of $2\times10^{-12}/\sqrt{\rm{Hz}}$. The two lasers pass through acoustic optical modulators (AOMs) before reaching the atoms. The RF signals driving the AOMs are derived from a stable quartz oscillator with stability of 25 ppm. The frequency of the RF radiation is locked to a GPS referenced Rb clock, which has the same frequency stability as the frequency comb. In total, the uncertainty in the determination of the excitation frequency is less than $2~\rm{kHz}$.

The spectral measurements were performed after the release of the atoms from optical molasses, so that there is no AC Stark shift due to the trapping fields. The excitation lasers and the RF field can cause AC Stark shifts. The AC Stark shift induced by the $\sim509$~nm laser is no greater than 400 Hz. The $\sim852$~nm laser induces a measured ($36.0\pm4.8$)~kHz blueshift of the ground state~\cite{Shen2024}. The 36.0 kHz shift is subtracted from the transition frequency and the uncertainty, 4.8 kHz, is taken as a contribution to the uncertainty of the resonance frequency.

\begin{figure}[!ht]
    \centering
    \includegraphics[width=0.48\textwidth]{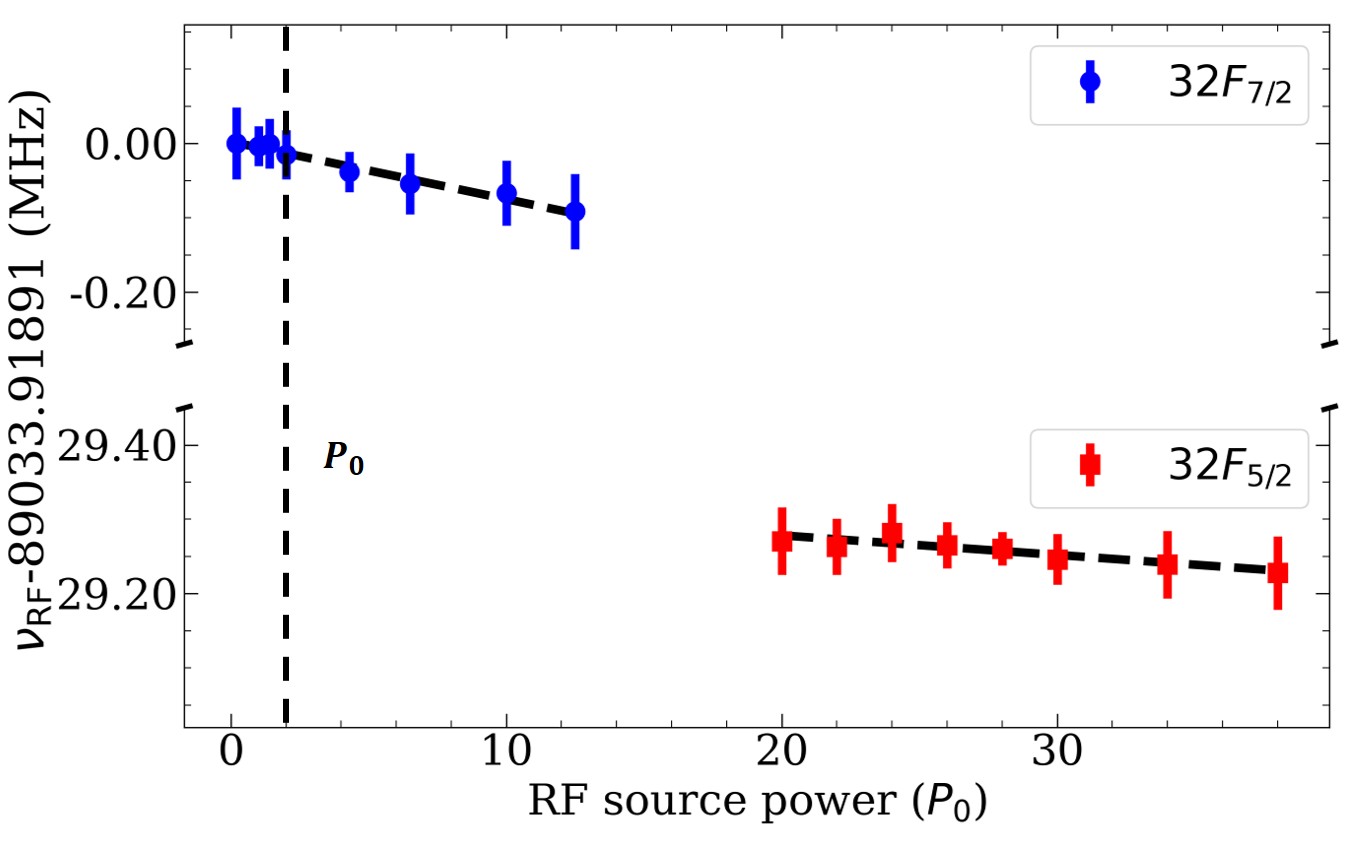}
    \caption{RF field induced AC Stark shift measurement. The absolute frequency of the $6S_{1/2} \rightarrow 32F_{5/2}$ (red squares) and $6S_{1/2} \rightarrow 32F_{7/2}$ (blue circles) transitions are measured at different power levels of the RF field. The power $P_0=-30$~dBm is the RF source power setting to obtain the spectrum of $32F_{7/2}$ Rydberg state, which is about 20 times less than the power used to obtain the $32F_{5/2}$ spectrum. Each data point represents the center frequency extracted from the Lorentzian fit of the spectrum and the error bar represents the fitting uncertainty. The data are fit to a linear function (black dashed line), yielding the intercept at zero RF power. }
    \label{fig:ACstarkshift}
\end{figure}

To characterize the AC Stark shift induced by the RF field, we measured the transition frequency from the $|6S_{1/2}, F=3\rangle$ state to the $32F_{5/2,7/2}$ state as a function of RF power. The results are shown in Fig.~\ref{fig:ACstarkshift}. A~linear fit is applied to extract the intercept for zero RF power. To obtain the spectrum of the $32F_{7/2}$ Rydberg state, the RF power is set to $P_0=-30$~dBm resulting in a $\sim (5.9\pm9.1)$ kHz red AC Stark shift. The power to obtain the $32F_{5/2}$ spectrum is about 20 times stronger than the power used to measure the $32F_{7/2}$ spectrum. This introduces a $(67.9\pm26.3)$~kHz red AC Stark shift to the energy level. During the transition frequency measurements, we kept the Rabi frequency of the three-photon transition ($|6S_{1/2}, F=3\rangle \rightarrow nF_{J}$) similar to the Rabi frequency used in the AC Stark shift measurements. Thus, we applied the measured AC Stark shift corrections (5.9~kHz for $nF_{7/2}$ and 67.9~kHz for $nF_{5/2}$) to all Rydberg states. The uncertainty of the measured AC Stark shift (9.1 kHz for $nF_{7/2}$ and 26.3 kHz for $nF_{5/2}$) contributes to the uncertainty of the absolute frequency measurement.

We minimize DC Stark shifts by compensating the stray electric fields via the eight electrodes around the trapped atoms, as shown in Fig.~\ref{fig:expill}(b). We use the $160S_{1/2}$ Rydberg state for calibrating and apply the same method described in Ref.~\cite{Shen2024,Merkt2013}. The background electric field that affects the atomic sample is reduced to $<1~\mathrm{mV}\,\mathrm{cm}^{-1}$. For the $nF_J$ series, the DC Stark effect causes inhomogeneous broadening more than a frequency shift because it leads to splitting of the $m_J$ sub-states. Based on the polarizability of Cs($nF_J$)~\cite{ARC2017}, the DC Stark effect induced uncertainty for the $68F_{7/2}$ state is no greater than $25~\mathrm{kHz}$, while for the $68F_{5/2}$ state, the uncertainty is no larger than $10~\rm{kHz}$.

Stray magnetic fields cause Zeeman shifts and splittings of the transitions. The stray magnetic field is compensated using three pairs of magnetic coils placed around the MOT chamber. Microwave spectroscopy is used to determine the amplitude of the total remaining magnetic field during the atom interrogation period~\cite{Shen2024}. The background magnetic field strength is reduced to $< 2$ mG during the measurement of the Rydberg state transition frequencies, which results in a maximum Zeeman splitting of around 14.3 kHz. The sensitivity to the magnetic field is similar across all Rydberg states because they have the same Land\'e $g$ factor. This maximum Zeeman splitting is treated as a common uncertainty for all Rydberg states.

The energy level shift due to Rydberg--Rydberg interaction is negligible. We tuned the 509~nm laser 20 MHz away from the resonance to adiabatically eliminate the presence of the $nD_{5/2}$ state atoms. The experiment was run under conditions where only 1 to 2 ions were created during each excitation and ionization cycle. The pressure shift due to collisions between the Rydberg atoms and the surrounding ground state atoms is a source of uncertainty. Using the theory in Ref.~\cite{Feng2010}, we estimated the collisional cross section, $\sigma=5.067\times10^{-11}~\rm cm^{-2}$ at $n=68$. With a peak Cs cloud density of $1\times 10^{10}~\rm cm^{-3}$, the maximum pressure shift is estimated to be $<6.0$~kHz for $n=68$. The main sources of uncertainty are summarized in Table~\ref{tab:uncertainty}.

\begin{table}[ht]
\caption{Uncertainties in the determination of the absolute frequency. The value of each uncertainty is listed as the maximum possible value for any states measured in this work.}
\label{tab:uncertainty} 
\begin{center}
\begin{tabular}{c|c}
\hline\hline
Source                       & Amplitude (kHz)  \\ \hline
Ground state AC Stark shift  & \multicolumn{1}{d{2.1}}{<\dd 5.0} \\  
Rydberg state AC Stark shift & \multicolumn{1}{d{2.1}}{<   26.3} \\  
Zeeman effect shift          & \multicolumn{1}{d{2.1}}{<   14.3} \\  
DC Stark shift               & \multicolumn{1}{d{2.1}}{<   25.0} \\  
Hyperfine splitting          & \multicolumn{1}{d{2.1}}{<\dd 4.0} \\  
Pressure shift               & \multicolumn{1}{d{2.1}}{<\dd 6.0} \\  
Statistical uncertainty      & \multicolumn{1}{d{2.1}}{<   24.0} \\  
\hline
Total                        & \multicolumn{1}{d{2.1}}{<   50.0} \\  
\hline\hline
\end{tabular}
\end{center}
\end{table}

\section{Ionization energy and quantum defects determination}

\begin{figure}[!ht]
    \centering
    \includegraphics[width=0.48\textwidth]{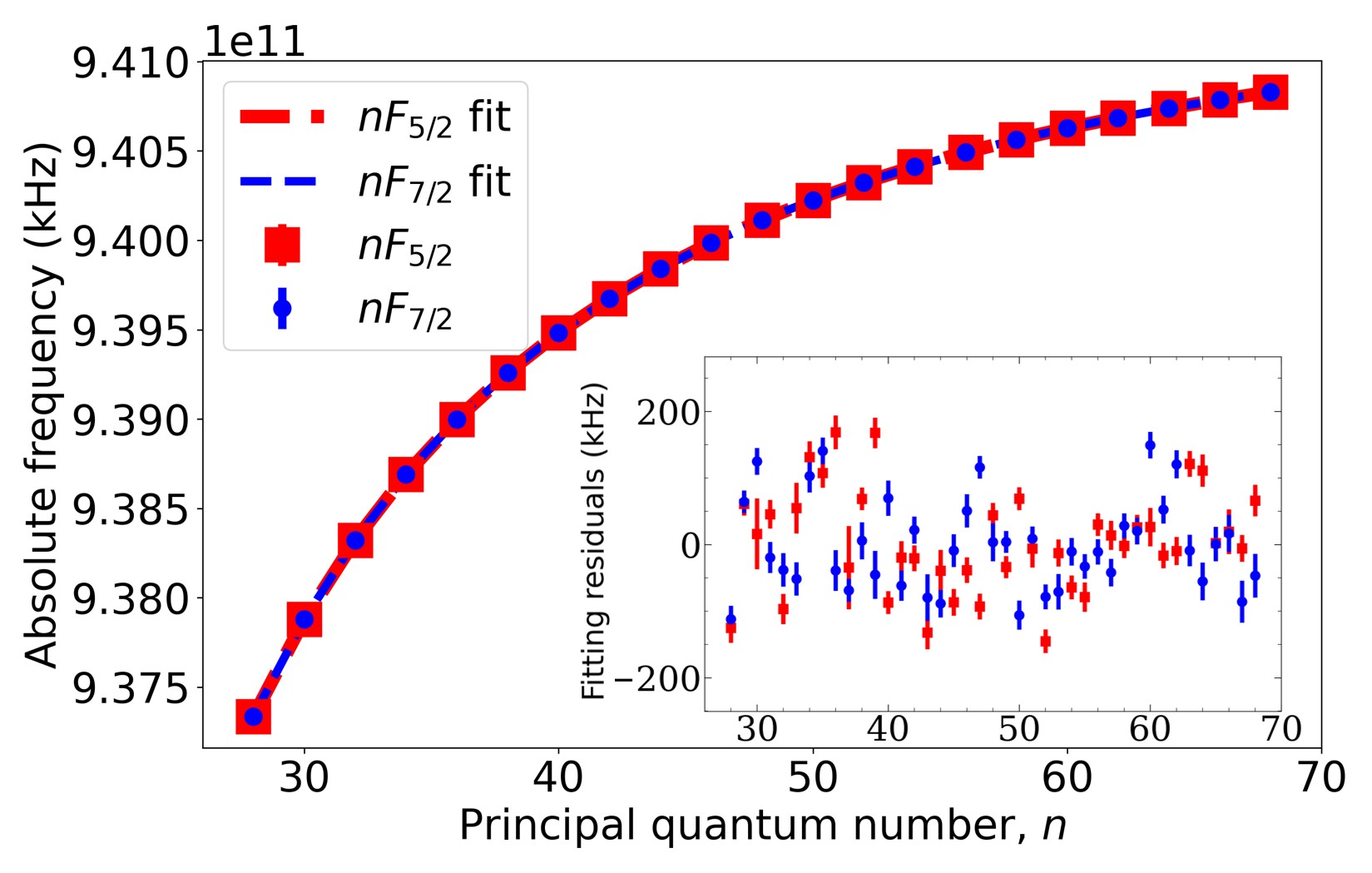}
    \caption{Absolute frequency measurements for the transitions from the $|6S_{1/2}, F=3\rangle$ hyperfine ground state to the $nF_{5/2}$ and $nF_{7/2}$ states. Data are presented with a step of $\Delta n=2$ to avoid crowding. The dashed lines are the modified Ritz formula fitting functions. The fitting residuals are given in the inset. The error bar represents the sum of the statistical uncertainty and the systematic uncertainty. The statistical uncertainties of the fitting residuals of the $nF_{5/2}$ and $nF_{7/2}$ series are 59.6 and 56.6~kHz, respectively.}
    \label{fig:absfrequency}
\end{figure}

The experimental term energies of a~Rydberg series, i.e., the excitation energies with respect to the ground state, are given by the modified Ritz formula:
\begin{equation}\label{Eq:QDequation}
E_{n,l,J} =  E_I - \frac{R}{[n-\delta^{(l,J)}(n)]^2},
\end{equation} 
where $E_I$ is the ionization energy and $R$ is the reduced mass Rydberg constant, which is equal to $R_{\infty}/(1+m_{\rm e}/m_{\rm core})$. Based on the  mass of the Cs atom from Ref.~\cite{CIAAW}, we determined $R_{^{133}\rm Cs} = $ 109736.8627304~cm$^{-1}$ ($3289828381.115$~MHz)~\cite{Shen2024}. For an unperturbed series, the $(n,l,J)$-dependent quantum defects are
\begin{equation}\label{eq:QDexpan}
\delta^{(l,J)}(n) = \delta^{(l,J)}_0 + \sum_{k=1}^{\infty} \frac{\delta^{(l,J)}_{{2k}}}{[n-\delta^{(l,J)}_0]^{2k}}.
\end{equation}
Using the nonlinear least squares method, we fitted the measured energy levels of the $nF_{J}$ series to Eq.~(\ref{eq:QDexpan}) to extract the ionization energy and the quantum defects. The residuals from the fits are uniformly distributed around zero, see Fig.~\ref{fig:absfrequency} and its inset. The values of the extracted ionization energies and the quantum defects are presented in Table~\ref{tab:quantdef}. The ionization energy from each series, $31406.467 751 52(25) \rm~cm^{-1}$ from the $nF_{5/2}$ series and $31406.467 751 46(26) \rm~cm^{-1}$ from the $nF_{7/2}$ series, agrees with our previous reported value, $31 406.467 751 48 (14) \rm~cm^{-1}$, which is the uncertainty-weighted average across all $nS_{1/2}$, $nD_{3/2}$, and $nD_{5/2}$ series \cite{Shen2024}.

\begin{figure}[!ht]
    \centering
    \includegraphics[width=0.48\textwidth]{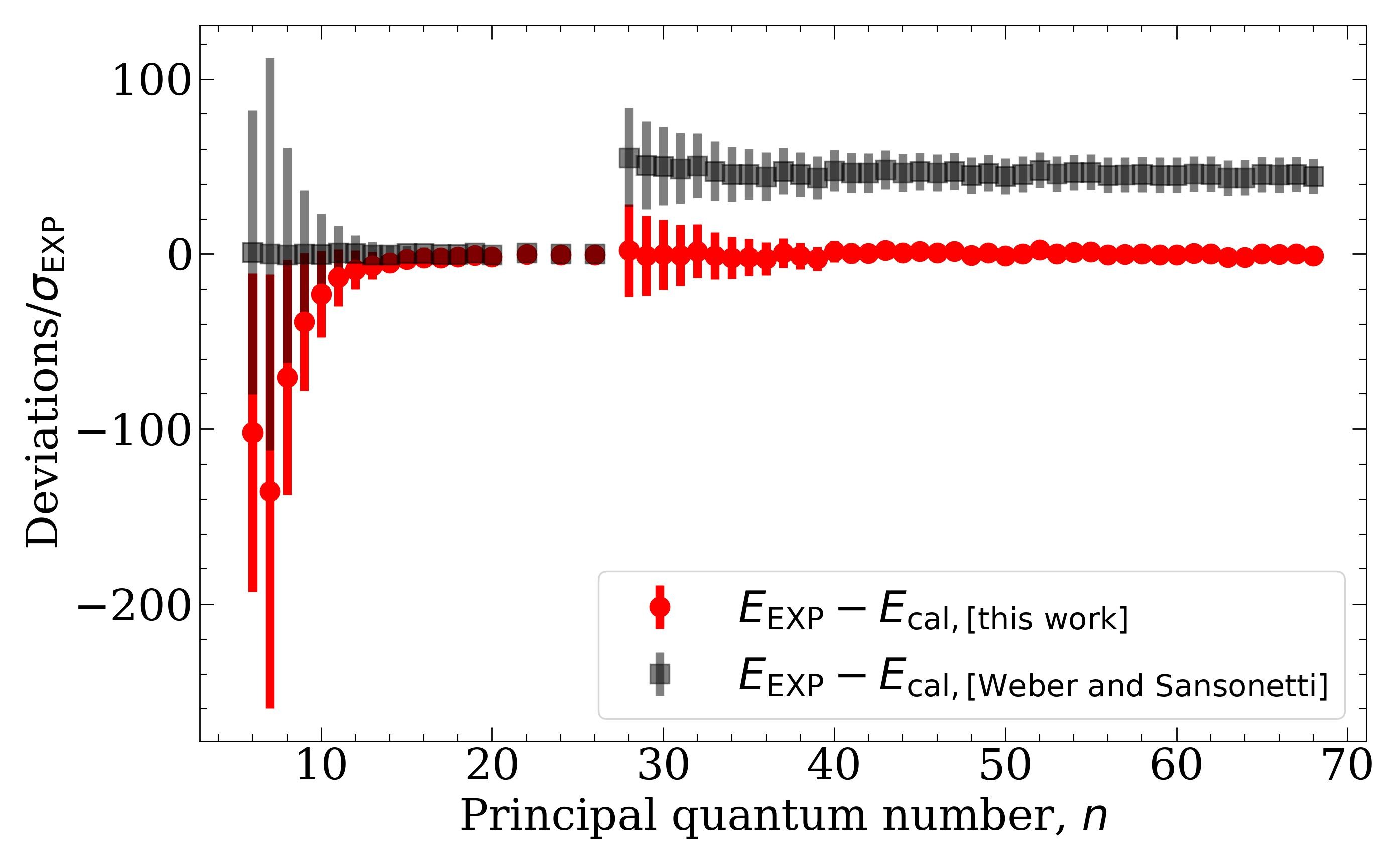}
    \caption{Deviations characterized by the experimental uncertainty ($\sigma_{\rm EXP}$) between the observed energy levels, $E_{\rm EXP}$,~\cite{Shen2024,Weber1987} and the theoretical calculations with the modified Ritz formula, $E_{\rm cal}$. Black squares and red circles represent the deviations$/\sigma_{\rm EXP}$ between the experimental results and the theoretical calculations using the $nF_{5/2}$ quantum defects in Ref.~\cite{Weber1987} and this work, respectively. The error bars are the square root of the square sum of the propagated uncertainties from the uncertainties in the quantum defects and the experimental uncertainty.}
    \label{fig:deviations}
\end{figure}

\begin{table*}[!ht]
    \centering
        \caption{Quantum defects for the $nF_{5/2}$ and $nF_{7/2}$ series obtained by fitting the experimentally measured absolute frequencies of the Rydberg energy levels to Eq.~(\ref{Eq:QDequation}). The quoted uncertainties are the estimated standard deviations from the fit. As a comparison, the quantum defects from previous works \cite{Bai2023,Weber1987,Allinson2025} are also listed in this table. $E_I$ stands for the ionization energy in cm$^{-1}$.}
    \scalebox{0.95}{    
    \begin{tabular}{@{}c@{}|c|c|c|c|c|c|c@{}}
    \hline
    \hline
       & $nF_{5/2}$~[this work] & $nF_{5/2}$~\cite{Weber1987} & $nF_{5/2}$~\cite{Bai2023} &$nF_{5/2}$~\cite{Allinson2025}& $nF_{7/2}$~[this work] & $nF_{7/2}$~\cite{Bai2023}&$nF_{7/2}$~\cite{Allinson2025} \\ 
                        &($n=28$--68)          & ($n=4$--65)          &
                        ($n=45$--50)       &                        ($n=15$--19)         & 
                        ($n=28$--68)         & ($n=45$--50) &     
                        ($n=15$--19)
                        \\ \hline     
    $E_I$ & 31 406.467 751 52(25)& 31 406.467 66(2)     & ---                & ---            & 31 406.467 751 46(26)& ---                & ---             \\
       $\delta_0^{(l,J)}$ &      0.033 414 93(18)&      0.033 414 24(96)&    0.033 415 37(70)&    0.033 429(3)&      0.033 562 89(19)&    0.033 5646(13)  &    0.033 570(3) \\ 
       $\delta_2^{(l,J)}$ &   $-$0.200 36(14)    &   $-$0.198 674       & $-$0.2014(16)      & $-$0.2025(12)  &   $-$0.202 89(14)    & $-$0.2052(29)      & $-$0.2016(11)   \\
       $\delta_4^{(l,J)}$ &      0.2825(8)       &      0.289 53        &    0.0             &    0.69(13)    &      0.2998(9)       &    0.0             &    0.45(11)     \\
       $\delta_6^{(l,J)}$ &      0.0             &   $-$0.2601          &    0.0             &    0.0         &      0.0             &    0.0             &    0.0          \\
       \hline \hline
    \end{tabular}
    }
    \label{tab:quantdef}
\end{table*}

\begin{table*}[!ht]
    \centering
        \caption{A summary of Cs quantum defects for the $nS_{1/2}$, $nP_J$, $nD_J$, and $nF_J$ series. The ionization energy value is taken as $31406.46775148(14)~\rm cm^{-1}$~\cite{Shen2024}.
        }
    \begin{tabular}{@{}c@{}|c|c|c|c|c|c|c@{}}
    \hline
    \hline
       &$nS_{1/2}$~\cite{Shen2024} & $nP_{1/2}$~\cite{Deiglmayr2016} & $nP_{3/2}$~\cite{Deiglmayr2016} & $nD_{3/2}$~\cite{Shen2024} & $nD_{5/2}$~\cite{Shen2024} & $nF_{5/2}$~[this work] & $nF_{7/2}$~[this work] \\ 
       &($n=21$--90)      & ($n=27$--74)  & ($n=27$--74)  &($n=23$--90) & ($n=23$--90) &($n=28$--68)  &($n=28$--68) \\ \hline
       $\delta_0^{(l,J)}$ & 4.049 359 94(17)& 3.591 5871(3) & 3.559 0676(3) &   2.475 458 40(13)&   2.466 315 29(14)&   0.033 414 93(18)&   0.033 562 89(19)\\          
       $\delta_2^{(l,J)}$ & 0.238 017(18)   & 0.362 73(16)  & 0.374 69(14)  &   0.008 339(62)   &   0.013 431(65)   &$-$0.200 36(14)    &$-$0.202 89(14)    \\         
       $\delta_4^{(l,J)}$ & 0.1747(33)      & 0.0           & 0.0           &$-$0.3769(95)      &$-$0.3613(61)      &   0.2825(8)       &   0.2998(9)       \\
       \hline \hline
    \end{tabular}
    \label{tab:quantdefCs}
\end{table*}

For comparison, we list in Table~\ref{tab:quantdef} the quantum defects for the $nF_{J}$ series published in Ref.~\cite{Weber1987,Bai2023,Allinson2025}. In Ref.~\cite{Weber1987}, the energy levels were determined by measuring laser wavelengths with a~high-precision Fabry-Perot interferometer. The uncertainty was 6 MHz. In our work, we measure the energy levels with narrow linewidth lasers and an RF field yielding less than 60 kHz uncertainty. Bai {\it et al.}~\cite{Bai2023} also used RF spectroscopy to extract the energy intervals between $nD_{5/2}$ and $(n-2)F_{J}$ Rydberg levels with a comparable uncertainty to our work. However, they depend on knowing the quantum defects of the $nD_{5/2}$ series, taken from Ref.~\cite{Weber1987}, to extract the quantum defects of the $nF_{J}$ series. The uncertainty reported in Ref.~\cite{Bai2023} is limited by the uncertainty in Ref.~\cite{Weber1987}. The work presented in this paper improves the uncertainty by two orders of magnitude.

We report the quantum defects up to the third term in the expansion~(\ref{eq:QDexpan}), where $k=2$, for both the $nF_{5/2} (n=28$--$68)$ and $nF_{7/2} (n=28$--$68)$ states, while Weber and Sansonetti provided them up to the fourth-order term to describe the energy levels for the $nF_{5/2}(n=4$--$65)$ states~\cite{Weber1987}. We compare the energy levels obtained from our quantum defects and those of Weber and Sansonetti~\cite{Weber1987} with the observed energy levels for $nF_{5/2}(n=6$--$68)$. The experimental data includes the measurements for $nF_{5/2}(n=6$--$27)$ in Ref.~\cite{Weber1987} and the measurements for $nF_{5/2}(n=28$--$68)$ in this work. We calculate the deviations characterized by the experimental uncertainty between the observed data and theoretical calculations using different sets of the quantum defects, as shown in Fig.~\ref{fig:deviations}. We find that the prior quantum defects with the nonzero expansion $k=3$ term, obtained by fitting to observed energies for low-$n$ Rydberg levels, do not reproduce our measurements for higher-$n$ levels within uncertainty. However, the quantum defect parameters reported in this work for high-$n$ levels, where $\delta^{(l,J)}_{2k}=0$ for $k=3$,  reproduce the previous observed levels at low~$n$. Our measurements show that the higher expansion $k>2$ terms are not necessary to reproduce all the available data, including the low-lying states, within the experimental uncertainty for all of the measurements. This is consistent with our previous conclusion that quantum defects for $nS_{1/2}$ and $nD_J$ with up to $k=2$ terms can reproduce the energy levels for low~$n$~\cite{Shen2024}.

With the inclusion of the prior measurements for the $nP_J$ series~\cite{Merkt2013} and the most recent precision measurements for the $nS_{1/2}$ and $nD_J$ series~\cite{Shen2024}, we construct a~database for quantum defects of Cs, as shown in Table~\ref{tab:quantdefCs}.

\section{Evaluation of the core penetration and core polarization contributions}\label{sec_Numerov}

\begin{figure}[!t]
    \centering
    \includegraphics[width=0.48\textwidth]{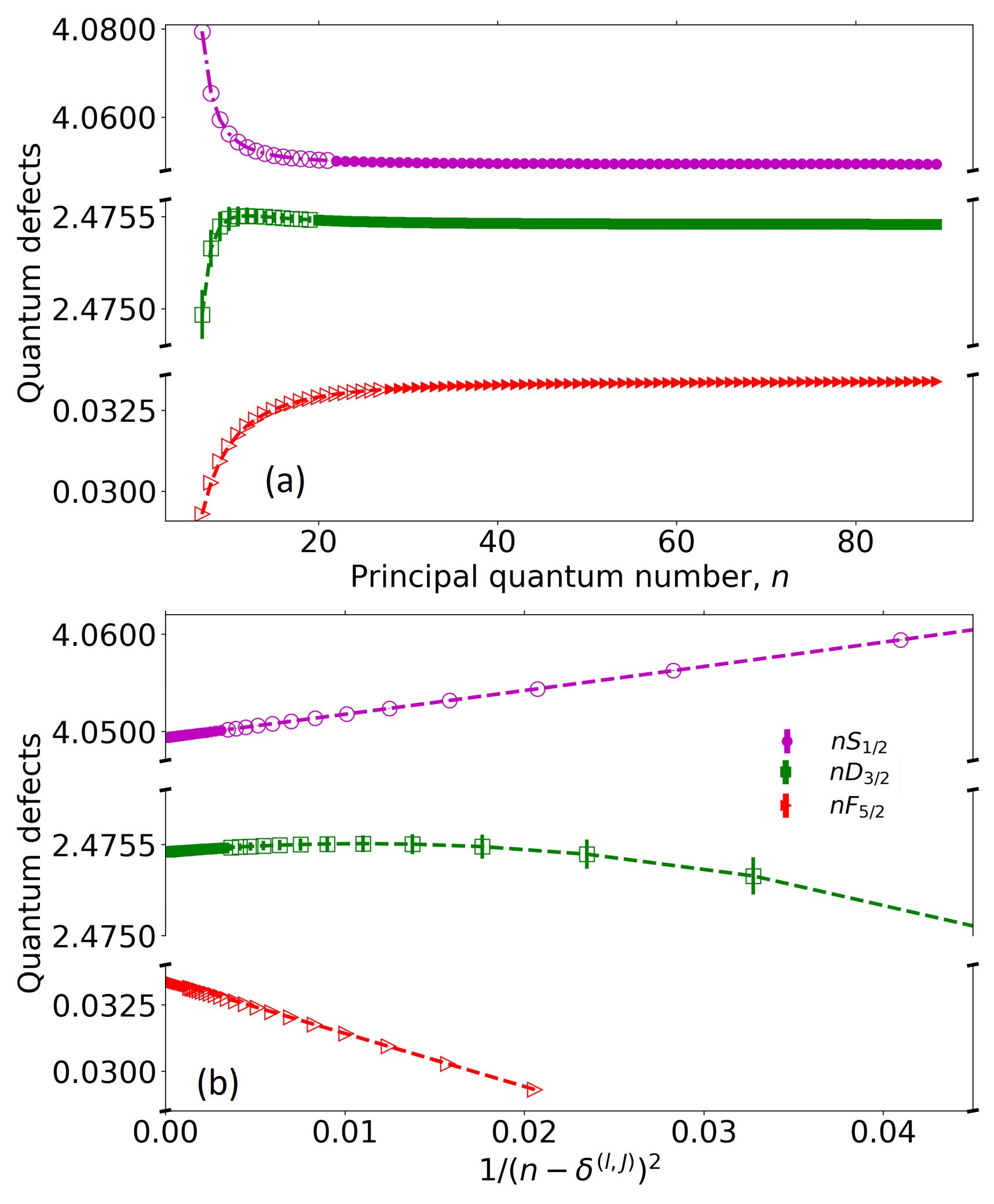}
    \caption{Experimental quantum defects $\delta^{(l,J)}$ versus (a) the principal quantum number $n$ and (b) $1/[n-\delta^{(l,J)}]^2$, for the $nF_{5/2}$ series and the previously measured $nS_{1/2}$ and $nD_{3/2}$ series~\cite{Shen2024} in Cs. 
    Solid markers are the measured quantum defects, while open markers are the fitting extrapolations from the experimental measurements. 
    The error bar is acquired by propagating the fitting uncertainty of the quantum defects.}
    \label{fig:QDvsN}
\end{figure}

We plot the quantum defects as a function of principal quantum number for the $nS_{1/2}$, $nD_{3/2}$ and $nF_{5/2}$ series, as shown in Fig.~\ref{fig:QDvsN}(a). The trend of the $nD_J$ series is nonmonotonic; the quantum defect initially increases with respect to $n$, reaches a peak and then decreases to an asymptote. In turn, the quantum defect for $nS_{1/2}$ decreases monotonically and for $nF_{J}$ increases monotonically with $n$. The behavior can be viewed more clearly by plotting the quantum defects as a function of $1/[n-\delta^{(l,J)}]^2$, Fig.~\ref{fig:QDvsN}(b). As studied in our previous work~\cite{Shen2024}, the behavior of the $nD_J$ series is the result of a competition between core polarization and core penetration. Core polarization reduces the quantum defect as the electron effective radius is reduced, i.e., low~$n$. As $n$ increases, the core polarization effect decreases faster than the core penetration effect, resulting in a quantum defect dominated by core penetration. In addition, the core penetration effect decreases as the angular momentum, $l$, increases due to the change of the outer electron trajectory. For the $nS_{1/2}$ series, the core penetration effect is dominant, while for the $nF_{J}$ series, the core polarization effect is dominant.

To evaluate the core penetration and core polarization contributions, we consider the Hamiltonian for the outermost electron of an alkali-metal atom in the form (in a.u.)
\begin{equation} \label{Hamil}
\hat H = -\frac{1}{2\mu}\nabla^2 + V_l(r) + V_l^{\rm LS}(r),     
\end{equation}
where $\mu$ is the reduced mass, $V_l(r)$ is the model one-electron  potential~\cite{Marinescu1994}, and $V_l^{\rm LS}(r)$ is the spin--orbit coupling~\cite{Aymar_1996}. 
We adopt the $l$-dependent parametric potential developed by Marinescu \textit{et al.}~\cite{Marinescu1994} 
\begin{eqnarray}
 V_l(r) & = & -\frac{1}{r} -\frac{(Z-1)}{r} \exp \left [-a_1^{(l)} r \right ] \nonumber \\ 
 & & + \left  (a_3^{(l)}+a_4^{(l)} r \right) \exp \left [-a_2^{(l)} r \right ]  \nonumber \\
 & & - \frac{\alpha_c}{2r^4}\left ( 1-\exp \left [-\left (r/r_c^{(l)} \right )^6 \right ] \right )
 \label{eq:modelV}
\end{eqnarray}
with $Z$ being the nuclear charge number of the neutral atom and $\alpha_c$ being the static dipole polarizability of the core. The terms in the first and second lines of Eq.~(\ref{eq:modelV}) can be viewed as the penetration potential $(V_{\rm{pen}})$, whereas the third-line term can be viewed as the polarization potential $(V_{\rm{pol}})$. The nontrivial form of the spin--orbit interaction reads 
(in a.u.)~\cite{Aymar_1996}
\begin{equation} \label{potVLS}
V^{\rm LS}_l(r) =  {\textbf{L}}\cdot{\textbf{S}}  \frac{1}{2\mu^2 c^2} \frac{1}{r}\frac{{\rm d}V_l(r)}{{\rm d}r}    \left (1 -\frac{1}{2\mu^2 c^2}V_l(r)  \right )^{-2},
\end{equation}
where $\langle {\textbf{L}}\cdot{\textbf{S}} \rangle^{(l,J)} =  (\hbar^2/2)[J(J+1)-l(l+1)-3/4]$ and $c$ is the speed of light. The last part of $V^{\rm LS}_l$ ensures proper behavior of the radial wave functions at very short range.

We compute the wave functions of the Schr\"odinger equation for the quantum defect energies by numerical integration using the Numerov method. The integration interval was taken as $[0.2, 2n(n+15)]$ (in a.u.) and divided into $4\times10^6$ subintervals.
We calculated the expectation values of $V_{\rm{pen}}$ and $V_{\rm{pol}}$, allowing us to estimate the core penetration and core polarization contributions to $\delta_0^{(l,J)}$. These contributions are presented in Table~\ref{tab:QDcontributions}. For the $nD_J$ series, the contribution from core polarization is slightly smaller than the contribution from core penetration, while for the $nF_{J}$ series, the contribution from core polarization is two orders of magnitude higher than the contribution from core penetration. In turn, for the $nS_{1/2}$ series, the contribution from core penetration is a factor of 30 higher than the contribution from core polarization.

\begin{table}
    \centering
        \caption{Numerically computed core polarization contribution, $\delta_{\rm pol}^{(l,J)}$, and core penetration contribution, $\delta_{\rm pen}^{(l,J)}$, to the leading quantum defect expansion term of the $nS_{1/2}$, $nD_{3/2}$, $nD_{5/2}$, $nF_{5/2}$, and $nF_{7/2}$ series.}
    \begin{tabular}{l|l|l|l}
    \hline
    \hline
         \multicolumn{1}{c|}{Series}  & \multicolumn{1}{c|}{$\delta_0^{(l,J)}$}  & \multicolumn{1}{c|}{$\delta_{\rm pol}^{(l,J)}$} & \multicolumn{1}{c}{$\delta_{\rm pen}^{(l,J)}$}  \\  \hline
         $nS_{1/2}$~\cite{Shen2024} & 4.04935994(17)  & 0.1320(5)   &\dl 3.9111(4)   \\
         $nD_{3/2}$~\cite{Shen2024} & 2.47545840(13)  & 1.0625(7)   &\dl 1.4094(5)  \\         
         $nD_{5/2}$~\cite{Shen2024} & 2.46631529(14)  & 1.0482(4)   &\dl 1.4197(6)  \\ 
         $nF_{5/2}$~[this work]     & 0.03341493(18)  & 0.033787(1) &$-$0.000394(6) \\
         $nF_{7/2}$~[this work]     & 0.03356289(19)  & 0.034065(2) &$-$0.000526(6) \\
         \hline \hline
    \end{tabular}
    \label{tab:QDcontributions}
\end{table}

\section{Fine structure intervals}

\begin{table}[!b]
    \centering
        \caption{Fine-structure interval parameters determined for $nF_J$ series from this work and other previous works~\cite{Bai2023,Goy_PRA_1982,Fredriksson1980}. The quoted uncertainties are the standard deviations of the fit.} 
    \begin{tabular}{@{}l@{}|@{}l@{}|l@{}|@{}l@{}}    
    \hline \hline
      $nF_{5/2} \rightarrow nF_{7/2}$ & \multicolumn{1}{c|}{$\xi^{(l)}_1$ (kHz) } & \multicolumn{1}{c|}{$\xi^{(l)}_2$ (kHz)} & \multicolumn{1}{c}{$\xi^{(l)}_3$ (kHz)} \\ \hline 
      Exp [this work]                         & $-$9.812(26)$\times10^{8}$ & 3.4(6)$\times10^{10}$   & $-$9.8(3.8)$\times10^{12}$   \\  
      Exp \cite{Bai2023}                      & $-$9.785(61)$\times10^{8}$ & 1.8(1.3)$\times10^{10}$ & \multicolumn{1}{c}{$0$}    \\          
      Exp \cite{Goy_PRA_1982,Fredriksson1980} & $-$9.796$\times10^{8}$     & 1.222$\times10^{10}$    & $-$3.376$\times10^{10}$ \\         
      \hline\hline
    \end{tabular}
    \label{tab:finespara}
\end{table}

\begin{figure}[!ht]
    \centering
    \includegraphics[width=0.48\textwidth]{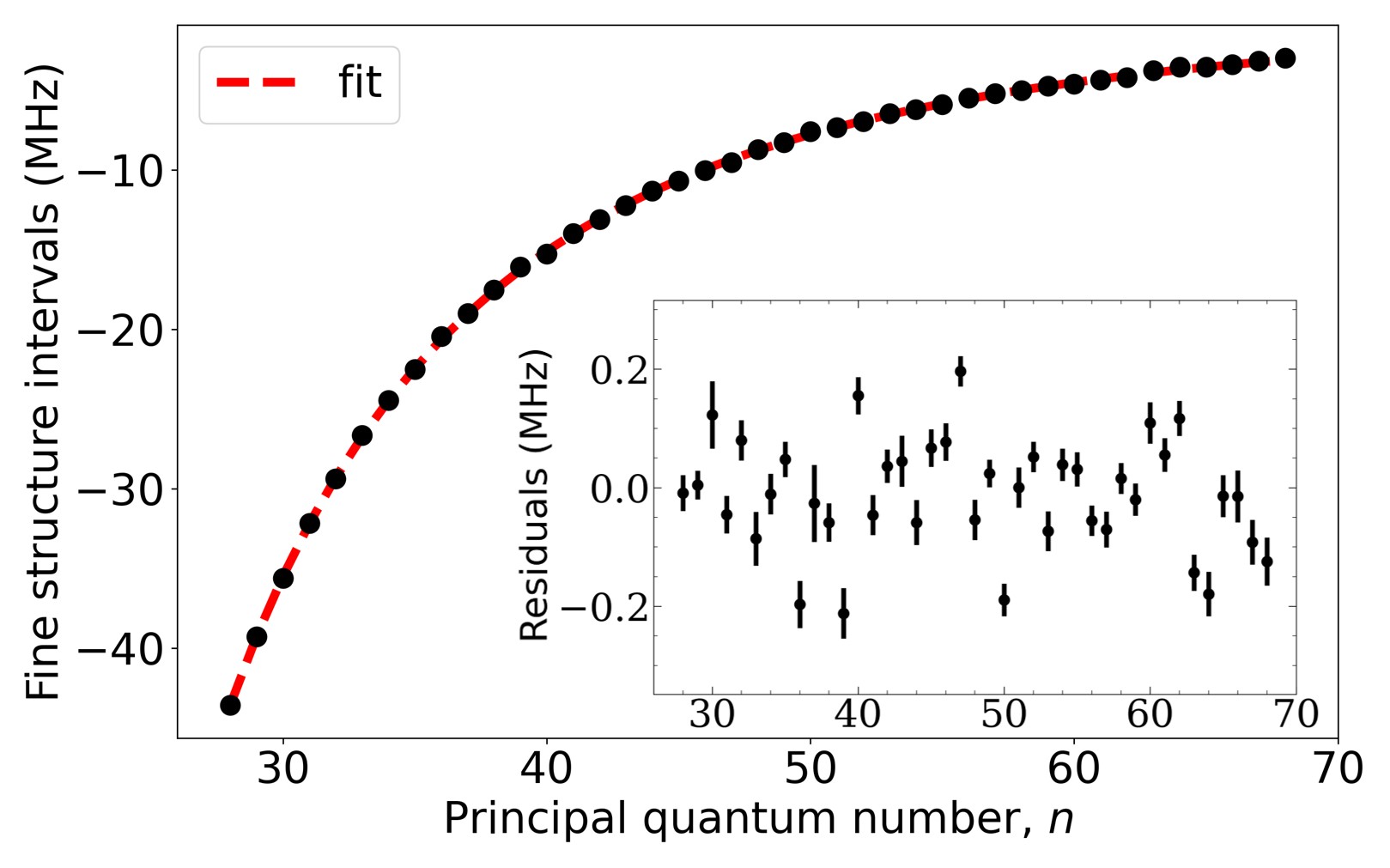}
    \caption{The fine-structure frequency separation for the $nF_J$ series. The red dashed line represents the fitting function given by Eq.~(\ref{eq:finestrucsep}). The fitting residuals are given as the black circles in the inset.}
    \label{fig:finestructuresep}
\end{figure}

The fine-structure splitting of the $nF_J$ levels can be expressed as an expansion in inverse odd powers of $[n-\tilde{\delta}^{(l,J=l\pm1/2)}]$ as follows~\cite{Goy_PRA_1982}
\begin{equation}\label{eq:finestrucsep}
\Delta_{\rm fs}^{(l)}(n) = \sum_{q=1}^{\infty} \frac{\xi^{(l)}_q}{[n-\tilde{\delta}^{(l,J=l\pm1/2)}(n)]^{2q+1}},
\end{equation}
where $\tilde{\delta}^{(l,J=l\pm1/2)}(n)$ is the weighted average quantum defect given by~\cite{Shen2024} 
\begin{equation}\label{eq:averagedelta}
\tilde{\delta}^{(l,J=l\pm1/2)}(n) = w^{(l,J<l)} {\delta}^{(l,J<l)}(n) + w^{(l,J>l)} {\delta}^{(l,J>l)}(n).
\end{equation}
The weights $w^{(l,J)}$ are equal to $1-|\langle {\textbf{L}}\cdot{\textbf{S}} \rangle^{(l,J)}| / [\hbar^2(l+1/2)]$. As $l$ increases, the weight tends to 1/2. For the $nF_J$ series, $w^{(3,5/2)}=3/7$ and $w^{(3,7/2)}=4/7$. Our prior work~\cite{Shen2024} revealed that the expansion~(\ref{eq:finestrucsep}) can be analytically justified in calculating $ \langle {\textbf{L}}\cdot{\textbf{S}} (1/r)({\rm d} V/{\rm d}r) \rangle$ for the hydrogen eigenfunctions and the Coulomb potential along with the polarization term.  

Using the values in Table~\ref{tab:quantdef} and representing $\delta^{(l,J)}$ in Eq.~(\ref{eq:averagedelta}) by the dominant term of the expansion~(\ref{eq:QDexpan}), we fitted our measured fine-structure intervals to Eq.~(\ref{eq:finestrucsep}) and extracted the $\xi^{(l)}_q$ parameters up to $q=3$. In Table~\ref{tab:finespara}, we compare our parameters with those of other authors reported in Ref.~\cite{Goy_PRA_1982,Fredriksson1980} and recently in Ref.~\cite{Bai2023}. Our uncertainty has been improved by a~factor of 2 due to the wider range of measurements ($n=28$--$68$) and smaller experimental uncertainty ($<60~\rm kHz$). The fitting residuals are randomly distributed around zero as shown in the inset of Fig.~\ref{fig:finestructuresep}. The average quantum defect $\tilde{\delta}^{(l,J=l\pm1/2)}(n)$ should be weighted by the $l$ and $J$ quantum numbers of the $nF_{5/2}$ and $nF_{7/2}$ series, rather than assigning them equal weight as done in previous studies~\cite{Goy_PRA_1982, Fredriksson1980, Bai2023}.

\section{Reduced dipole matrix elements}

\begin{table}[h]
    \centering
        \caption{The reduced electric-dipole matrix elements (in a.u.). \label{tab_red_mat} }
    \begin{tabular}{c|r|r}
    \hline
    \hline
        Transition & \multicolumn{1}{c|}{Others}  & \multicolumn{1}{c}{This work}  \\ \hline
        $7D_{3/2} \leftrightarrow 5F_{5/2}   $ &  43.4(2)~\cite{Safronova2016}   &  43.4028  \\
        $7D_{5/2} \leftrightarrow 5F_{5/2}   $ &  11.66(5)~\cite{Safronova2016}  &  11.6573  \\
        $7D_{5/2} \leftrightarrow 5F_{7/2}   $ &  52.2(2)~\cite{Safronova2016}   &  52.1325  \\ \hline        
        $8D_{3/2} \leftrightarrow 6F_{5/2}   $ &  65.2(5)~\cite{Safronova2016}   &  65.1963  \\
        $8D_{5/2} \leftrightarrow 6F_{5/2}   $ &  17.5(1)~\cite{Safronova2016}   &  17.5317  \\
        $8D_{5/2} \leftrightarrow 6F_{7/2}   $ &  78.4(6)~\cite{Safronova2016}   &  78.4055  \\ \hline        
        $9D_{3/2} \leftrightarrow 7F_{5/2}   $ &  90.5(9)~\cite{Safronova2016}   &  90.4715  \\
        $9D_{5/2} \leftrightarrow 7F_{5/2}   $ &  24.4(2)~\cite{Safronova2016}   &  24.3460  \\
        $9D_{5/2} \leftrightarrow 7F_{7/2}   $ &  108.9(0)~\cite{Safronova2016}  & 108.8832  \\ \hline
        $10D_{3/2} \leftrightarrow 6F_{5/2}  $ &  \multicolumn{1}{c|}{---}       &   7.9896  \\
        $10D_{5/2} \leftrightarrow 6F_{5/2}  $ &  2.14(2)~\cite{Safronova2016}   &   2.1428  \\
        $10D_{5/2} \leftrightarrow 6F_{7/2}  $ &  9.56(8)~\cite{Safronova2016}   &   9.5828  \\ \hline
        $10D_{3/2} \leftrightarrow 8F_{5/2}  $ &  \multicolumn{1}{c|}{---}       & 119.3401  \\
        $10D_{5/2} \leftrightarrow 8F_{5/2}  $ &  32.1(3)~\cite{Safronova2016}   &  32.1304  \\
        $10D_{5/2} \leftrightarrow 8F_{7/2}  $ &  143(1)~\cite{Safronova2016}    & 143.6999  \\ \hline
        $11D_{3/2} \leftrightarrow 7F_{5/2}  $ &  \multicolumn{1}{c|}{---}       &  11.5014  \\
        $11D_{5/2} \leftrightarrow 7F_{5/2}  $ &  3.08(3)~\cite{Safronova2016}   &   3.0874  \\
        $11D_{5/2} \leftrightarrow 7F_{7/2}  $ &  13.8(1)~\cite{Safronova2016}   &  13.8073  \\ \hline
        $12D_{3/2} \leftrightarrow 8F_{5/2}  $ &  \multicolumn{1}{c|}{---}       &  15.5252  \\
        $12D_{5/2} \leftrightarrow 8F_{5/2}  $ &  4.21(4)~\cite{Safronova2016}   &   4.1706  \\
        $12D_{5/2} \leftrightarrow 8F_{7/2}  $ &  18.8(2)~\cite{Safronova2016}   &  18.6519  \\ \hline        
        $30D_{3/2} \leftrightarrow 30F_{5/2} $ &   47.8433~\cite{ARC2017}   &  47.8384   \\
        $30D_{5/2} \leftrightarrow 30F_{5/2} $ &   11.8521~\cite{ARC2017}   &  11.8508   \\
        $30D_{5/2} \leftrightarrow 30F_{7/2} $ &   53.0042~\cite{ARC2017}   &  52.9305   \\ \hline
        $40D_{3/2} \leftrightarrow 40F_{5/2} $ &   92.4414~\cite{ARC2017}   &  92.4387   \\
        $40D_{5/2} \leftrightarrow 40F_{5/2} $ &   23.0681~\cite{ARC2017}   &  23.0672   \\
        $40D_{5/2} \leftrightarrow 40F_{7/2} $ &  103.1637~\cite{ARC2017}   & 103.0400   \\\hline
        $50D_{3/2} \leftrightarrow 50F_{5/2} $ &  150.9921~\cite{ARC2017}   & 150.9931   \\
        $50D_{5/2} \leftrightarrow 50F_{5/2} $ &   37.8184~\cite{ARC2017}   &  37.8183   \\
        $50D_{5/2} \leftrightarrow 50F_{7/2} $ &  169.1289~\cite{ARC2017}   & 168.9428   \\ \hline
        $60D_{3/2} \leftrightarrow 60F_{5/2} $ &  223.5003~\cite{ARC2017}   & 223.5067   \\
        $60D_{5/2} \leftrightarrow 60F_{5/2} $ &   56.1042~\cite{ARC2017}   &  56.1053   \\ 
        $60D_{5/2} \leftrightarrow 60F_{7/2} $ &  250.9056~\cite{ARC2017}   & 250.6445   \\ \hline
        $70D_{3/2} \leftrightarrow 70F_{5/2} $ &   309.9680~\cite{ARC2017}  & 309.9814   \\
        $70D_{5/2} \leftrightarrow 70F_{5/2} $ &    77.9261~\cite{ARC2017}  &  77.9288   \\
        $70D_{5/2} \leftrightarrow 70F_{7/2} $ &   348.4960~\cite{ARC2017}  & 348.1474   \\ \hline
        $80D_{3/2} \leftrightarrow 80F_{5/2} $ &   410.3962~\cite{ARC2017}  & 410.4183   \\
        $80D_{5/2} \leftrightarrow 80F_{5/2} $ &   103.2843~\cite{ARC2017}  & 103.2890   \\
        $80D_{5/2} \leftrightarrow 80F_{7/2} $ &   461.9012~\cite{ARC2017}  & 461.4526   \\ \hline
        $30D_{3/2} \leftrightarrow 60F_{5/2} $ &     0.4544~\cite{ARC2017}  &  0.4575    \\
        $30D_{5/2} \leftrightarrow 60F_{5/2} $ &     0.0941~\cite{ARC2017}  &  0.0947    \\
        $30D_{5/2} \leftrightarrow 60F_{7/2} $ &     0.4209~\cite{ARC2017}  &  0.4215    \\ \hline
        $30D_{3/2} \leftrightarrow 80F_{5/2} $ &     0.2176~\cite{ARC2017}  &  0.2196    \\
        $30D_{5/2} \leftrightarrow 80F_{5/2} $ &     0.0431~\cite{ARC2017}  &  0.0437    \\
        $30D_{5/2} \leftrightarrow 80F_{7/2} $ &     0.1929~\cite{ARC2017}  &  0.1942    \\
    \hline \hline
    \end{tabular}
\end{table}

\begin{table*}[t!] 
\caption {\label{tab:polariz} The convergence of the scalar and tensor polarizabilities (in a.u.)\ for the selected Rydberg states with respect to $n^\prime$ in Eqs.~(\ref{eqn:scalar}) and~(\ref{eqn:tensor}). The notation $\{x\}$ denotes $\times 10^x$. The closed interval $[a,b]$ has the form $\{n' \in \mathbb{N} \, | \, a \le n' \le b \}$. }
\centering
\begin{tabular}{cc|llll|lll}
\hline\hline
$n$ &  $n^\prime$   & \multicolumn{1}{c}{$\alpha_0(S_{1/2})$}  & 
                      \multicolumn{1}{c}{$\alpha_0(P_{3/2})$}  & 
                      \multicolumn{1}{c}{$\alpha_0(D_{5/2})$}  & 
                      \multicolumn{1}{c|}{$\alpha_0(F_{7/2})$} & 
                      \multicolumn{1}{c}{$\alpha_2(P_{3/2})$}  & 
                      \multicolumn{1}{c}{$\alpha_2(D_{5/2})$}  &
                      \multicolumn{1}{c}{$\alpha_2(F_{7/2})$} \\
\hline
10 & $\le 120$ & 4.765478\{5\} & 4.532621\{6\} & $-$5.329947\{6\} & 2.848115\{8\} & $-$4.517002\{5\} & 6.875773\{6\} & $-$1.157954\{8\} \\
   & $\le 100$ & 4.765476\{5\} & 4.532619\{6\} & $-$5.329948\{6\} & 2.848115\{8\} & $-$4.516998\{5\} & 6.875774\{6\} & $-$1.157953\{8\} \\
   & $\le  80$ & 4.765474\{5\} & 4.532616\{6\} & $-$5.329950\{6\} & 2.848114\{8\} & $-$4.516990\{5\} & 6.875776\{6\} & $-$1.157953\{8\} \\
   & $\le  60$ & 4.765468\{5\} & 4.532608\{6\} & $-$5.329955\{6\} & 2.848113\{8\} & $-$4.516973\{5\} & 6.875779\{6\} & $-$1.157953\{8\} \\
   & $\le  40$ & 4.765450\{5\} & 4.532583\{6\} & $-$5.329971\{6\} & 2.848110\{8\} & $-$4.516918\{5\} & 6.875791\{6\} & $-$1.157951\{8\} \\
   & $\le  20$ & 4.765287\{5\} & 4.532388\{6\} & $-$5.330109\{6\} & 2.848076\{8\} & $-$4.516438\{5\} & 6.875905\{6\} & $-$1.157933\{8\} \\
   & $\le  15$ & 4.764873\{5\} & 4.531985\{6\} & $-$5.330533\{6\} & 2.847954\{8\} & $-$4.515227\{5\} & 6.876301\{6\} & $-$1.157852\{8\} \\
\hline 
30 & $\le 120$ & 5.112055\{9\} & 9.715287\{10\} & $-$6.363966\{10\} & 6.888778\{11\} & $-$8.651025\{9\} & 7.146096\{10\} & $-$2.810909\{11\} \\
   & $\le 100$ & 5.112054\{9\} & 9.715286\{10\} & $-$6.363967\{10\} & 6.888778\{11\} & $-$8.651023\{9\} & 7.146096\{10\} & $-$2.810909\{11\} \\
   & $\le  80$ & 5.112052\{9\} & 9.715285\{10\} & $-$6.363967\{10\} & 6.888777\{11\} & $-$8.651021\{9\} & 7.146097\{10\} & $-$2.810909\{11\} \\
   & $\le  60$ & 5.112044\{9\} & 9.715282\{10\} & $-$6.363969\{10\} & 6.888776\{11\} & $-$8.651012\{9\} & 7.146099\{10\} & $-$2.810908\{11\} \\
   & $\le  40$ & 5.111950\{9\} & 9.715252\{10\} & $-$6.364000\{10\} & 6.888762\{11\} & $-$8.650912\{9\} & 7.146129\{10\} & $-$2.810899\{11\} \\
   & $\le  35$ & 5.111585\{9\} & 9.715151\{10\} & $-$6.364183\{10\} & 6.888661\{11\} & $-$8.650466\{9\} & 7.146310\{10\} & $-$2.810816\{11\} \\
   &$[10,120]$ & 5.112055\{9\} & 9.715287\{10\} & $-$6.363966\{10\} & 6.888778\{11\} & $-$8.651025\{9\} & 7.146096\{10\} & $-$2.810909\{11\} \\
   &$[20,120]$ & 5.112060\{9\} & 9.715288\{10\} & $-$6.363964\{10\} & 6.888778\{11\} & $-$8.651026\{9\} & 7.146095\{10\} & $-$2.810909\{11\} \\
   &$[25,120]$ & 5.112250\{9\} & 9.715336\{10\} & $-$6.363771\{10\} & 6.888780\{11\} & $-$8.651092\{9\} & 7.146036\{10\} & $-$2.810910\{11\} \\
\hline
60 & $\le 120$ & 7.936442\{11\} & 1.947755\{13\} & $-$1.150465\{13\} & 8.888329\{13\} & $-$1.699182\{12\} & 1.263127\{13\} & $-$3.629749\{13\} \\
   & $\le 100$ & 7.936439\{11\} & 1.947755\{13\} & $-$1.150465\{13\} & 8.888329\{13\} & $-$1.699182\{12\} & 1.263127\{13\} & $-$3.629749\{13\} \\
   & $\le  80$ & 7.936421\{11\} & 1.947755\{13\} & $-$1.150465\{13\} & 8.888328\{13\} & $-$1.699181\{12\} & 1.263127\{13\} & $-$3.629748\{13\} \\
   & $\le  70$ & 7.936323\{11\} & 1.947753\{13\} & $-$1.150468\{13\} & 8.888319\{13\} & $-$1.699172\{12\} & 1.263130\{13\} & $-$3.629743\{13\} \\
   & $\le  65$ & 7.935764\{11\} & 1.947741\{13\} & $-$1.150495\{13\} & 8.888223\{13\} & $-$1.699113\{12\} & 1.263156\{13\} & $-$3.629660\{13\} \\
   &$[10,120]$ & 7.936442\{11\} & 1.947755\{13\} & $-$1.150465\{13\} & 8.888329\{13\} & $-$1.699182\{12\} & 1.263127\{13\} & $-$3.629749\{13\} \\
   &$[30,120]$ & 7.936442\{11\} & 1.947755\{13\} & $-$1.150465\{13\} & 8.888329\{13\} & $-$1.699182\{12\} & 1.263127\{13\} & $-$3.629749\{13\} \\
   &$[50,120]$ & 7.936487\{11\} & 1.947756\{13\} & $-$1.150463\{13\} & 8.888330\{13\} & $-$1.699184\{12\} & 1.263126\{13\} & $-$3.629750\{13\} \\
   &$[55,120]$ & 7.937206\{11\} & 1.947775\{13\} & $-$1.150416\{13\} & 8.888337\{13\} & $-$1.699211\{12\} & 1.263111\{13\} & $-$3.629753\{13\} \\
\hline\hline
\end{tabular}
\end{table*}

Using the energies from our quantum defects and their corresponding wave functions for $nD_{J}$ and $nF_{J}$ series, we compute reduced electric-dipole matrix elements for $nD_J \leftrightarrow n'F_{J'}$ transitions of Cs as follows
\begin{eqnarray}\label{eqn:dipole}
\langle J||r||J' \rangle & = & (-1)^{l+J'+s+1} \sqrt{(2J+1)(2J'+1)} \delta_{s,s'}   \nonumber \\
 & & \times 
\begin{Bmatrix}
 l && J && s \\
 J' && l' && 1 \\
\end{Bmatrix} \langle l||r||l' \rangle,   
\end{eqnarray}
where the curly brackets $\{ : : : \}$ denote the Wigner $6j$ symbol and 
\begin{eqnarray}
\langle l||r||l' \rangle &=&  (-1)^{l} \sqrt{(2l+1)(2l'+1)}   \nonumber \\
& & \times 
\begin{pmatrix}
 l && 1 && l' \\
 0 && 0 && 0 \\
\end{pmatrix}
\langle nlJ |r| n'l'J' \rangle. 
\end{eqnarray} 
The results are presented in Table~\ref{tab_red_mat}. For  transitions between low excited states, the reduced dipole moments can be  compared to the relativistic all-order many-body calculations of Safronova {\it et al.}~\cite{Safronova2016}. Excellent agreement is obtained within theoretical uncertainties. It should be emphasized that by utilizing our quantum defects, determined from high-lying levels, one can obtain reliable wave functions. Different matrix elements across all $n$, including for low $n$ excited states can be accurately calculated. In turn, the reduced electric-dipole matrix elements for high-$n$ Rydberg states in Table~\ref{tab_red_mat} are contrasted with those computed with the ARC library~\cite{ARC2017}. Since ARC uses the $nF_{5/2}$ quantum defects of Weber and Sansonetti~\cite{Weber1987} for calculating both the $nF_{5/2}$ and $nF_{7/2}$ series, our results show good agreement with calculations based on the ARC library for $nF_{5/2}$, matching up to the third decimal place, and significantly less agreement for $nF_{7/2}$. This emphasizes the vital role of our quantum defects in ensuring high-quality calculations.




\section{Scalar and tensor polarizabilities}

We calculate the scalar and tensor polarizabilities, $\alpha_0$ and $\alpha_2$, respectively, of Cs$(n,l,J)$ states according to~\cite{Khadjavi_1968, Lai_polarizabilities}
\begin{eqnarray} 
  \alpha_0 & = &  -\frac{2}{3} \sum_{n',l',J'} (2J'+1)   
     \begin{Bmatrix} 
        l & J & 1/2  \\
        J' & l' & 1  \\
     \end{Bmatrix}^2  \nonumber \\
    & & \times \text{max}(l, l') \frac{ | \langle nlJ |r | n'l'J' \rangle |^2}{E_{n,l,J}-E_{n',l',J'}},  \label{eqn:scalar} 
\end{eqnarray}
\begin{eqnarray}    
  \alpha_2 & = & -2 \left[ \frac{10J(2J-1)(2J+1)}{3(J+1)(2J+3)} \right]^{1/2}    \nonumber \\
 & &
  \times   \sum_{n',l',J'} (-1)^{J+J'}(2J'+1) 
        \text{max}(l, l')    \begin{Bmatrix} 
   l & J & 1/2  \\
   J' & l' & 1  \\
   \end{Bmatrix}^2 \nonumber \\
 & & \times      
   \begin{Bmatrix} 
   J & J' & 1  \\
   1 & 2 & J  \\
   \end{Bmatrix}
   \frac{ | \langle nlJ |r | n'l'J' \rangle |^2}{E_{n,l,J}-E_{n',l',J'}}. \label{eqn:tensor}  
\end{eqnarray}
Our polarizability calculations are highly reliable and consistent. Table~\ref{tab:polariz} illustrates how the polarizabilities converge as the number of terms in the above equations increases  for specific states. It is clear that the most significant contributions to the polarizabilities come from the nearest intermediate states. To perform polarizability calculations for the $nF_J$ states, one must have proper energies and wave functions for the $l\pm1$ states, i.e., for the $nD_J$ and $nG_J$ series. Therefore, to find the $nG_J$ states, we calculated the energies from quantum defects given in Ref.~\cite{Allinson2025} and then determined the corresponding eigenfunctions based on our technique discussed in Sec.~\ref{sec_Numerov}. Convergence of all obtained results is observed with increasing range of $n'$.

\begin{table}[!b] 
\caption {\label{tab:coeff} Expansion coefficients for the scalar $(\alpha_0)$ and tensor $(\alpha_2)$ polarizabilities of Cs Rydberg states in a.u., see Eq.~(\ref{eqn:coeff}).}
\centering
\begin{tabular}{@{}cd{4.11}@{}d{4.10}@{}d{4.8}@{}}
\hline\hline
$\alpha_{i}^{(n,l,J)}$  &  \multicolumn{1}{c}{$\eta_{i,0}^{(l,J)}$}   &   \multicolumn{1}{c}{$\eta_{i,1}^{(l,J)}$}  &  \multicolumn{1}{c}{$\eta_{i,2}^{(l,J)}$}    \\
\hline
$\alpha_{0} (nS_{1/2})$  &  0.30544462(54)   & 8.752230(89)    & 1.572(4)      \\ 
$\alpha_{0} (nP_{1/2})$  &  8.0161209(16)    & 8.75278(27)     & -60.935(11)   \\
$\alpha_{0} (nP_{3/2})$  & 10.5476819(21)    & 8.74918(34)     & -87.169(14)   \\
$\alpha_{0} (nD_{3/2})$  & -4.7608174(17)    & 8.7710(3)       & 32.167(12)    \\
$\alpha_{0} (nD_{5/2})$  & -5.678563(2)      & 8.77181(31)     & 42.087(13)    \\
$\alpha_{0} (nF_{5/2})$  & 32.069148(8)      & 8.8311(14)      & -338.71(6)    \\
$\alpha_{0} (nF_{7/2})$  & 31.820387(9)      & 8.8372(16)      & -331.09(7)    \\
\hline
$\alpha_{2} (nP_{3/2})$  &  -0.90277020(45)  & -1.75135(8)     & 7.9828(31)    \\
$\alpha_{2} (nD_{3/2})$  &   3.0740604(7)    & -1.75681(12)    & -20.4264(46)  \\
$\alpha_{2} (nD_{5/2})$  &   6.111517(1)     & -2.51143(17)    & -48.232(7)    \\
$\alpha_{2} (nF_{5/2})$  & -11.2733688(25)   & -2.53002(43)    & 118.69(2)     \\
$\alpha_{2} (nF_{7/2})$  & -13.0053439(35)   & -2.9546(6)      & 134.711(26)   \\
\hline\hline
\end{tabular}
\end{table}

Tables~\ref{tab_polar_scalar} and \ref{tab_polar_tensor} present the calculated scalar and tensor polarizabilities, respectively, of the selected cesium $nS_{1/2}$, $nP_{J}$, $nD_{J}$, and $nF_{J}$ states up to $n = 100$. Note that $\alpha_2(S_{1/2})$ and $\alpha_2(P_{1/2})$ are always equal to zero. The results are compared with some available experimental~\cite{Zhao_2011, Bai2020, Lei_1995} and theoretical~\cite{Safronova2016, Wijngaarden_1994, Hou_2025, Yerokhin_2016, Bai_2024, Bhowmik_2014} findings. Moreover, the polrizabilites from the newer version of the ARC package (ver. 3.0)~\cite{ARCv3_2020} are also given in the tables. For low excited states, which our measurements did not reach, one can see a~good agreement between the results, especially when one pays attention to the calculations performed with the sophisticated relativistic method used by Safronova and coworkers~\cite{Safronova2016}. The vast majority of our values fall within their uncertainties. Interestingly, the results of van Wijngaarden and Li~\cite{Wijngaarden_1994}, obtained using a~simple method based on the Coulomb approximation, are similar to ours for the $nS_{1/2}$, $nP_{J}$, $nD_{J}$ states. However, their results for the $nF_J$ series completely differ from both our findings and those obtained by ARC, not only in magnitude but also in sign, which may indicate the inadequacy of the applied approximation in Ref.~\cite{Wijngaarden_1994}.

The theoretical values of Yerokhin {\it et al.}~\cite{Yerokhin_2016} and Hou {\it et al.}~\cite{Hou_2025} for higher lying levels are close to ours. Recently, Bai {\it et al.} measured the polarizabilities of Cs($nS_{1/2}$) Rydberg states using Stark spectroscopy~\cite{Bai2020}. Their results are considerably lower than our findings. Nevertheless, they have recently calculated $\alpha_0$ and $\alpha_2$ for the $37F_J$ and $39D_{5/2}$ states and validated these calculations with measurements, showing consistency within the reported uncertainties~\cite{Bai_2024}. An impressive agreement with our results can be observed; for example, we obtained $\alpha_0(37F_{5/2}) = 3.0244 \times 10^{12}$~a.u. and $\alpha_0(37F_{7/2}) = 3.0014 \times 10^{12}$~a.u., whereas Bai {\it et al.}~\cite{Bai_2024} reported $\alpha_0(37F_{5/2}) = 3.03 \times 10^{12}$~a.u. and $\alpha_0(37F_{7/2}) = 3.00 \times 10^{12}$~a.u, respectively. A~similar level of agreement is found for the tensor polarizability. It is worth emphasizing that ARC yields a~significantly less accurate value for $\alpha_0(37F_{7/2})=2.3930 \times 10^{12}$~a.u. Such an underestimation is also seen for other $\alpha_0(nF_{7/2})$. In turn, the ARC results for $\alpha_2$ of all the Rydberg $nF_{5/2}$ and $nF_{7/2}$ states appear to be inverted when compared with our findings and those of Ref.~\cite{Bai_2024}, as seen in the last two columns in Table~\ref{tab_polar_tensor}.

\begin{table*}[!h] 
\caption{\label{tab_polar_scalar} Selected scalar polarizabilities ($\alpha_0$) of the $nS_{1/2}$, $nP_J$, $nD_J$, and $nF_J$ states of Cs in a.u. The notation $\{x\}$ denotes $\times 10^x$.}
\centering
\begin{scriptsize}
\begin{tabular}{clllllll}
\hline\hline
$n$ & \multicolumn{1}{c}{$\alpha_0(S_{1/2})$} 
    & \multicolumn{1}{c}{$\alpha_0(P_{1/2})$}
    & \multicolumn{1}{c}{$\alpha_0(P_{3/2})$} 
    & \multicolumn{1}{c}{$\alpha_0(D_{3/2})$} 
    & \multicolumn{1}{c}{$\alpha_0(D_{5/2})$} 
    & \multicolumn{1}{c}{$\alpha_0(F_{5/2})$} 
    & \multicolumn{1}{c}{$\alpha_0(F_{7/2})$} \\ \hline
8 & 3.8122\{4\} & 2.2597\{5\} & 2.8824\{5\} & $-$3.7263\{5\} & $-$4.8011\{5\} & 5.5406\{7\} & 5.5190\{7\}  \\
 & 3.8256\{4\}~\cite{ARCv3_2020} & 2.2354\{5\}~\cite{ARCv3_2020} & 2.8418\{5\}~\cite{ARCv3_2020} & $-$3.7000\{5\}~\cite{ARCv3_2020} & $-$4.7699\{5\}~\cite{ARCv3_2020} & 5.5481\{7\}~\cite{ARCv3_2020} & 5.5105\{7\}~\cite{ARCv3_2020}  \\
 & 3.827(26)\{4\}~\cite{Safronova2016}  & 2.233(14)\{5\}~\cite{Safronova2016}  & 2.845(17)\{5\}~\cite{Safronova2016}  & $-$3.693(58)\{5\}~\cite{Safronova2016}  & $-$4.758(63)\{5\}~\cite{Safronova2016} &  &   \\
 & 3.79\{4\}~\cite{Wijngaarden_1994}  & 2.21\{5\}~\cite{Wijngaarden_1994} & 2.82\{5\}~\cite{Wijngaarden_1994} & $-$3.66\{5\}~\cite{Wijngaarden_1994} & $-$4.72\{5\}~\cite{Wijngaarden_1994} & $-$9.34\{5\}~\cite{Wijngaarden_1994} & $-$1.03\{6\}~\cite{Wijngaarden_1994}  \\ \hline
10 & 4.7655\{5\} & 3.5050\{6\} & 4.5326\{6\} & $-$4.2465\{6\} & $-$5.3299\{6\} & 2.8680\{8\} & 2.8481\{8\}  \\
 & 4.7868\{5\}~\cite{ARCv3_2020} & 3.5003\{6\}~\cite{ARCv3_2020} & 4.5236\{6\}~\cite{ARCv3_2020} & $-$4.2392\{6\}~\cite{ARCv3_2020} & $-$5.3463\{6\}~\cite{ARCv3_2020} & 2.8639\{8\}~\cite{ARCv3_2020} & 2.2882\{8\}~\cite{ARCv3_2020}  \\
 & 4.775(33)\{5\}~\cite{Safronova2016} & 3.500(14)\{6\}~\cite{Safronova2016} & 4.525(20)\{6\}~\cite{Safronova2016} & $-$4.242(60)\{6\}~\cite{Safronova2016} & $-$5.324(59)\{6\}~\cite{Safronova2016} &  &   \\
 & 4.75\{5\}~\cite{Wijngaarden_1994}  & 3.49\{6\}~\cite{Wijngaarden_1994} & 4.51\{6\}~\cite{Wijngaarden_1994} & $-$4.22\{6\}~\cite{Wijngaarden_1994} & $-$5.30\{6\}~\cite{Wijngaarden_1994} & $-$4.44\{6\}~\cite{Wijngaarden_1994} & $-$4.94\{6\}~\cite{Wijngaarden_1994}  \\
 & 4.77\{5\}~\cite{Yerokhin_2016} & 3.56\{6\}~\cite{Yerokhin_2016} & 4.63\{6\}~\cite{Yerokhin_2016} & $-$4.36\{6\}~\cite{Yerokhin_2016} & $-$5.48\{6\}~\cite{Yerokhin_2016} &  &   \\ \hline
12 & 2.8630\{6\} & 2.4323\{7\} & 3.1648\{7\} & $-$2.5029\{7\} & $-$3.1036\{7\} & 1.0670\{9\} & 1.0595\{9\}  \\
 & 2.8680\{6\}~\cite{ARCv3_2020} & 2.4325\{7\}~\cite{ARCv3_2020} & 3.1627\{7\}~\cite{ARCv3_2020} & $-$2.5061\{7\}~\cite{ARCv3_2020} & $-$3.1080\{7\}~\cite{ARCv3_2020} & 1.0660\{9\}~\cite{ARCv3_2020} & 8.4947\{8\}~\cite{ARCv3_2020}  \\
 & 2.868(21)\{6\}~\cite{Safronova2016}  & 2.431(10)\{7\}~\cite{Safronova2016}  & 3.1630(96)\{7\}~\cite{Safronova2016} & $-$2.5092(783)\{7\}~\cite{Safronova2016} & $-$3.1487(802)\{7\}~\cite{Safronova2016} &  &   \\
 & 2.84\{6\}~\cite{Wijngaarden_1994} & 2.44\{7\}~\cite{Wijngaarden_1994} & 3.16\{7\}~\cite{Wijngaarden_1994} & $-$2.51\{7\}~\cite{Wijngaarden_1994} & $-$3.11\{7\}~\cite{Wijngaarden_1994} &  &   \\ 
 & 2.86\{6\}~\cite{Yerokhin_2016} & 2.46\{7\}~\cite{Yerokhin_2016} & 3.21\{7\}~\cite{Yerokhin_2016} & $-$2.54\{7\}~\cite{Yerokhin_2016} & $-$3.15\{7\}~\cite{Yerokhin_2016} &  &   \\ \hline
14 & 1.1578\{7\} & 1.0966\{8\} & 1.4315\{8\} & $-$1.0191\{8\} & $-$1.2546\{8\} & 3.2062\{9\} & 3.1833\{9\}  \\
 & 1.1588\{7\}~\cite{ARCv3_2020} & 1.0977\{8\}~\cite{ARCv3_2020} & 1.4330\{8\}~\cite{ARCv3_2020} & $-$1.0201\{8\}~\cite{ARCv3_2020} & $-$1.2562\{8\}~\cite{ARCv3_2020} & 3.2042\{9\}~\cite{ARCv3_2020} & 2.5490\{9\}~\cite{ARCv3_2020}  \\
 & 1.16\{7\}~\cite{Yerokhin_2016} & 1.10\{8\}~\cite{Yerokhin_2016} & 1.44\{8\}~\cite{Yerokhin_2016} & $-$1.03\{8\}~\cite{Yerokhin_2016} & $-$1.27\{8\}~\cite{Yerokhin_2016} &  &   \\
 & 1.14\{7\}~\cite{Wijngaarden_1994} & 1.10\{8\}~\cite{Wijngaarden_1994} & 1.43\{8\}~\cite{Wijngaarden_1994} &  &  &  &   \\ \hline
16 & 3.6474\{7\} & 3.7680\{8\} & 4.9281\{8\} & $-$3.2656\{8\} & $-$4.0008\{8\} & 8.2733\{9\} & 8.2137\{9\}  \\
 & 3.6488\{7\}~\cite{ARCv3_2020} & 3.7708\{8\}~\cite{ARCv3_2020} & 4.9329\{8\}~\cite{ARCv3_2020} & $-$3.2680\{8\}~\cite{ARCv3_2020} & $-$4.0059\{8\}~\cite{ARCv3_2020} & 8.2691\{9\}~\cite{ARCv3_2020} & 6.5709\{9\}~\cite{ARCv3_2020}  \\
 & 3.65\{7\}~\cite{Yerokhin_2016} & 3.78\{8\}~\cite{Yerokhin_2016} & 4.96\{8\}~\cite{Yerokhin_2016} & $-$3.28\{8\}~\cite{Yerokhin_2016} & $-$4.03\{8\}~\cite{Yerokhin_2016} &  &   \\
 & 3.54\{7\}~\cite{Wijngaarden_1994} &  &  &  &  &  &   \\ \hline
20 & 2.2591\{8\} & 2.6643\{9\} & 3.4919\{9\} & $-$2.1110\{9\} & $-$2.5702\{9\} & 4.0049\{10\} & 3.9754\{10\}  \\
 & 2.2586\{8\}~\cite{ARCv3_2020} & 2.6653\{9\}~\cite{ARCv3_2020} & 3.4953\{9\}~\cite{ARCv3_2020} & $-$2.1122\{9\}~\cite{ARCv3_2020} & $-$2.5737\{9\}~\cite{ARCv3_2020} & 4.0030\{10\}~\cite{ARCv3_2020} & 3.1764\{10\}~\cite{ARCv3_2020}  \\ \hline
25 & 1.2874\{9\} & 1.7087\{10\} & 2.2423\{10\} & $-$1.2676\{10\} & $-$1.5362\{10\} & 1.9276\{11\} & 1.9132\{11\}  \\
 & 1.2865\{9\}~\cite{ARCv3_2020} & 1.7090\{10\}~\cite{ARCv3_2020} & 2.2447\{10\}~\cite{ARCv3_2020} & $-$1.2681\{10\}~\cite{ARCv3_2020} & $-$1.5385\{10\}~\cite{ARCv3_2020} & 1.9266\{11\}~\cite{ARCv3_2020} & 1.5272\{11\}~\cite{ARCv3_2020}  \\ \hline
30 & 5.1121\{9\} & 7.3984\{10\} & 9.7153\{10\} & $-$5.2663\{10\} & $-$6.3640\{10\} & 6.9412\{11\} & 6.8888\{11\}  \\
 & 5.1065\{9\}~\cite{ARCv3_2020} & 7.3993\{10\}~\cite{ARCv3_2020} & 9.7263\{10\}~\cite{ARCv3_2020} & $-$5.2683\{10\}~\cite{ARCv3_2020} & $-$6.3736\{10\}~\cite{ARCv3_2020} & 6.9375\{11\}~\cite{ARCv3_2020} & 5.4956\{11\}~\cite{ARCv3_2020}  \\ \hline
35 & 1.6048\{10\} & 2.4823\{11\} & 3.2610\{11\} & $-$1.7175\{11\} & $-$2.0714\{11\} & 2.0480\{12\} & 2.0325\{12\}  \\
 & 1.6027\{10\}~\cite{ARCv3_2020} & 2.4825\{11\}~\cite{ARCv3_2020} & 3.2648\{11\}~\cite{ARCv3_2020} & $-$1.7181\{11\}~\cite{ARCv3_2020} & $-$2.0745\{11\}~\cite{ARCv3_2020} & 2.0469\{12\}~\cite{ARCv3_2020} & 1.6207\{12\}~\cite{ARCv3_2020}  \\ \hline
37 & 2.4144\{10\} & 3.8196\{11\} & 5.0185\{11\} & $-$2.6192\{11\} & $-$3.1567\{11\} & 3.0244\{12\} & 3.0014\{12\}  \\
 & 2.4110\{10\}~\cite{ARCv3_2020} & 3.8199\{11\}~\cite{ARCv3_2020} & 5.0245\{11\}~\cite{ARCv3_2020} & $-$2.6201\{11\}~\cite{ARCv3_2020} & $-$3.1616\{11\}~\cite{ARCv3_2020} & 3.0227\{12\}~\cite{ARCv3_2020} & 2.3930\{12\}~\cite{ARCv3_2020}  \\
 & 2.41\{10\}~\cite{Yerokhin_2016} & 3.82\{11\}~\cite{Yerokhin_2016} & 5.02\{11\}~\cite{Yerokhin_2016} & $-$2.62\{11\}~\cite{Yerokhin_2016} & $-$3.15\{11\}~\cite{Yerokhin_2016} & 3.03\{12\}~\cite{Bai_2024} & 3.00\{12\}~\cite{Bai_2024}   \\ \hline
39 & 3.5493\{10\} & 5.7321\{11\} & 7.5321\{11\} & $-$3.8993\{11\} & $-$4.6969\{11\} & 4.3751\{12\} & 4.3418\{12\}  \\
 & 3.5440\{10\}~\cite{ARCv3_2020} & 5.7324\{11\}~\cite{ARCv3_2020} & 7.5411\{11\}~\cite{ARCv3_2020} & $-$3.9006\{11\}~\cite{ARCv3_2020} & $-$4.7042\{11\}~\cite{ARCv3_2020} & 4.3726\{12\}~\cite{ARCv3_2020} & 3.4613\{12\}~\cite{ARCv3_2020}  \\
 & 3.55\{10\}~\cite{Yerokhin_2016} & 5.73\{11\}~\cite{Yerokhin_2016} & 7.53\{11\}~\cite{Yerokhin_2016} & $-$3.90\{11\}~\cite{Yerokhin_2016} & $-$4.69\{11\}~\cite{Yerokhin_2016} &  &   \\
 &  &  &  & $-$4.54(16)\{11\}~\cite{Zhao_2011} & $-$4.86(16)\{11\}~\cite{Zhao_2011} &  &   \\
 &  &  &  &  & $-$4.70\{11\}~\cite{Bai_2024}  &  &   \\ \hline
40 & 4.2696\{10\} & 6.9625\{11\} & 9.1493\{11\} & $-$4.7188\{11\} & $-$5.6826\{11\} & 5.2251\{12\} & 5.1852\{12\}  \\
 & 4.2631\{10\}~\cite{ARCv3_2020} & 6.9628\{11\}~\cite{ARCv3_2020} & 9.1603\{11\}~\cite{ARCv3_2020} & $-$4.7205\{11\}~\cite{ARCv3_2020} & $-$5.6915\{11\}~\cite{ARCv3_2020} & 5.2221\{12\}~\cite{ARCv3_2020} & 4.1335\{12\}~\cite{ARCv3_2020}  \\ \hline
42 & 6.0901\{10\} & 1.0114\{12\} & 1.3292\{12\} & $-$6.8085\{11\} & $-$8.1951\{11\} & 7.3564\{12\} & 7.3002\{12\}  \\
 & 6.0804\{10\}~\cite{ARCv3_2020} & 1.0114\{12\}~\cite{ARCv3_2020} & 1.3308\{12\}~\cite{ARCv3_2020} & $-$6.8108\{11\}~\cite{ARCv3_2020} & $-$8.2080\{11\}~\cite{ARCv3_2020} & 7.3521\{12\}~\cite{ARCv3_2020} & 5.8189\{12\}~\cite{ARCv3_2020}  \\
 & 6.09\{10\}~\cite{Yerokhin_2016} & 1.01\{12\}~\cite{Yerokhin_2016} & 1.33\{12\}~\cite{Yerokhin_2016} & $-$6.80\{11\}~\cite{Yerokhin_2016} & $-$8.18\{11\}~\cite{Yerokhin_2016} & &  \\
 &  &  &  & $-$5.87(36)\{11\}~\cite{Zhao_2011} & $-$8.56(96)\{11\}~\cite{Zhao_2011} &  &   \\ \hline
45 & 1.0044\{11\} & 1.7100\{12\} & 2.2475\{12\} & $-$1.1408\{12\} & $-$1.3723\{12\} & 1.1932\{13\} & 1.1841\{13\}  \\
 & 1.0027\{11\}~\cite{ARCv3_2020} & 1.7100\{12\}~\cite{ARCv3_2020} & 2.2503\{12\}~\cite{ARCv3_2020} & $-$1.1412\{12\}~\cite{ARCv3_2020} & $-$1.3745\{12\}~\cite{ARCv3_2020} & 1.1925\{13\}~\cite{ARCv3_2020} & 9.4368\{12\}~\cite{ARCv3_2020}  \\ \hline
50 & 2.1483\{11\} & 3.7908\{12\} & 4.9831\{12\} & $-$2.4979\{12\} & $-$3.0021\{12\} & 2.4969\{13\} & 2.4778\{13\}  \\
 & 2.1444\{11\}~\cite{ARCv3_2020} & 3.7908\{12\}~\cite{ARCv3_2020} & 4.9893\{12\}~\cite{ARCv3_2020} & $-$2.4988\{12\}~\cite{ARCv3_2020} & $-$3.0069\{12\}~\cite{ARCv3_2020} & 2.4954\{13\}~\cite{ARCv3_2020} & 1.9744\{13\}~\cite{ARCv3_2020}  \\
 & 2.15\{11\}~\cite{Yerokhin_2016,Bhowmik_2014} & 3.79\{12\}~\cite{Yerokhin_2016,Bhowmik_2014} & 4.98\{12\}~\cite{Yerokhin_2016,Bhowmik_2014} & $-$2.49\{12\}~\cite{Yerokhin_2016,Bhowmik_2014} & $-$3.00\{12\}~\cite{Yerokhin_2016,Bhowmik_2014} &  &   \\
 &  &  &  & $-$2.06(6)\{12\}~\cite{Zhao_2011}  & $-$2.79(10)\{12\}~\cite{Zhao_2011} &  &   \\ \hline
55 & 4.2591\{11\} & 7.7464\{12\} & 1.0184\{13\} & $-$5.0537\{12\} & $-$6.0693\{12\} & 4.8690\{13\} & 4.8316\{13\}  \\
 & 4.2508\{11\}~\cite{ARCv3_2020} & 7.7463\{12\}~\cite{ARCv3_2020} & 1.0197\{13\}~\cite{ARCv3_2020} & $-$5.0553\{12\}~\cite{ARCv3_2020} & $-$6.0789\{12\}~\cite{ARCv3_2020} & 4.8660\{13\}~\cite{ARCv3_2020} & 3.8494\{13\}~\cite{ARCv3_2020}  \\ \hline
60 & 7.9364\{11\} & 1.4814\{13\} & 1.9478\{13\} & $-$9.5852\{12\} & $-$1.1505\{13\} & 8.9572\{13\} & 8.8883\{13\}  \\
 & 7.9203\{11\}~\cite{ARCv3_2020} & 1.4814\{13\}~\cite{ARCv3_2020} & 1.9502\{13\}~\cite{ARCv3_2020} & $-$9.5882\{12\}~\cite{ARCv3_2020} & $-$1.1523\{13\}~\cite{ARCv3_2020} & 8.9516\{13\}~\cite{ARCv3_2020} & 7.0806\{13\}~\cite{ARCv3_2020}  \\
 & 8.22\{11\}~\cite{Bhowmik_2014} & 1.46\{13\}~\cite{Bhowmik_2014} & 1.87\{13\}~\cite{Bhowmik_2014} & $-$9.36\{12\}~\cite{Bhowmik_2014} & $-$1.19\{13\}~\cite{Bhowmik_2014} &  &   \\ \hline
65 & 1.4046\{12\} & 2.6813\{13\} & 3.5257\{13\} & $-$1.7229\{13\} & $-$2.0669\{13\} & 1.5692\{14\} & 1.5571\{14\}  \\
 & 1.4016\{12\}~\cite{ARCv3_2020} & 2.6813\{13\}~\cite{ARCv3_2020} & 3.5301\{13\}~\cite{ARCv3_2020} & $-$1.7235\{13\}~\cite{ARCv3_2020} & $-$2.0702\{13\}~\cite{ARCv3_2020} & 1.5682\{14\}~\cite{ARCv3_2020} & 1.2403\{14\}~\cite{ARCv3_2020}  \\
 & 1.2107(242)\{12\}~\cite{Bai2020}  &  &  &  &  &  &   \\ \hline
70 & 2.3797\{12\} & 4.6331\{13\} & 6.0924\{13\} & $-$2.9596\{13\} & $-$3.5490\{13\} & 2.6369\{14\} & 2.6166\{14\}  \\
 & 2.3745\{12\}~\cite{ARCv3_2020} & 4.6329\{13\}~\cite{ARCv3_2020} & 6.1001\{13\}~\cite{ARCv3_2020} & $-$2.9605\{13\}~\cite{ARCv3_2020} & $-$3.5547\{13\}~\cite{ARCv3_2020} & 2.6352\{14\}~\cite{ARCv3_2020} & 2.0840\{14\}~\cite{ARCv3_2020}  \\
 & 1.91\{12\}~\cite{Bhowmik_2014} & 4.74\{13\}~\cite{Bhowmik_2014} & 5.91\{13\}~\cite{Bhowmik_2014} & $-$3.14\{13\}~\cite{Bhowmik_2014} & $-$3.61\{13\}~\cite{Bhowmik_2014}  &  &   \\
 & 1.8389(246)\{12\}~\cite{Bai2020}  &  &  &  &  &  &   \\ \hline
75 & 3.8839\{12\} & 7.6941\{13\} & 1.0118\{14\} & $-$4.8902\{13\} & $-$5.8619\{13\} & 4.2751\{14\} & 4.2421\{14\}  \\
 & 3.8751\{12\}~\cite{ARCv3_2020} & 7.6939\{13\}~\cite{ARCv3_2020} & 1.0131\{14\}~\cite{ARCv3_2020} & $-$4.8917\{13\}~\cite{ARCv3_2020} & $-$5.8713\{13\}~\cite{ARCv3_2020} & 4.2723\{14\}~\cite{ARCv3_2020} & 3.3784\{14\}~\cite{ARCv3_2020}  \\
 & 2.9229(160)\{12\}~\cite{Bai2020}  &  & 1.0131\{14\}~\cite{Hou_2025} &  &  &  &   \\ \hline
80 & 6.1369\{12\} & 1.2347\{14\} & 1.6237\{14\} & $-$7.8129\{13\} & $-$9.3623\{13\} & 6.7178\{14\} & 6.6659\{14\}  \\
 & 6.1228\{12\}~\cite{ARCv3_2020} & 1.2346\{14\}~\cite{ARCv3_2020} & 1.6258\{14\}~\cite{ARCv3_2020} & $-$7.8152\{13\}~\cite{ARCv3_2020} & $-$9.3773\{13\}~\cite{ARCv3_2020} & 6.7134\{14\}~\cite{ARCv3_2020} & 5.3084\{14\}~\cite{ARCv3_2020}  \\
 &  &  & 1.6273\{14\}~\cite{Hou_2025} &  &  &  &   \\ \hline
85 & 9.4261\{12\} & 1.9228\{14\} & 2.5288\{14\} & $-$1.2120\{14\} & $-$1.4520\{14\} & 1.0271\{15\} & 1.0191\{15\}  \\
 & 9.4039\{12\}~\cite{ARCv3_2020} & 1.9228\{14\}~\cite{ARCv3_2020} & 2.5320\{14\}~\cite{ARCv3_2020} & $-$1.2124\{14\}~\cite{ARCv3_2020} & $-$1.4543\{14\}~\cite{ARCv3_2020} & 1.0264\{15\}~\cite{ARCv3_2020} & 8.1153\{14\}~\cite{ARCv3_2020}  \\
 &  &  & 2.5467\{14\}~\cite{Hou_2025} &  &  &  &   \\ \hline
90 & 1.4121\{13\} & 2.9166\{14\} & 3.8359\{14\} & $-$1.8321\{14\} & $-$2.1943\{14\} & 1.5326\{15\} & 1.5207\{15\}  \\
 & 1.4087\{13\}~\cite{ARCv3_2020} & 2.9164\{14\}~\cite{ARCv3_2020} & 3.8407\{14\}~\cite{ARCv3_2020} & $-$1.8327\{14\}~\cite{ARCv3_2020} & $-$2.1978\{14\}~\cite{ARCv3_2020} & 1.5315\{15\}~\cite{ARCv3_2020} & 1.2109\{15\}~\cite{ARCv3_2020}  \\ \hline
95 & 2.0688\{13\} & 4.3215\{14\} & 5.6838\{14\} & $-$2.7064\{14\} & $-$3.2407\{14\} & 2.2379\{15\} & 2.2206\{15\}  \\
 & 2.0638\{13\}~\cite{ARCv3_2020} & 4.3213\{14\}~\cite{ARCv3_2020} & 5.6910\{14\}~\cite{ARCv3_2020} & $-$2.7072\{14\}~\cite{ARCv3_2020} & $-$3.2459\{14\}~\cite{ARCv3_2020} & 2.2364\{15\}~\cite{ARCv3_2020} & 1.7681\{15\}~\cite{ARCv3_2020}  \\ \hline
100 & 2.9713\{13\} & 6.2706\{14\} & 8.2475\{14\} & $-$3.9164\{14\} & $-$4.6886\{14\} & 3.2049\{15\} & 3.1801\{15\}  \\
 & 2.9639\{13\}~\cite{ARCv3_2020} & 6.2703\{14\}~\cite{ARCv3_2020} & 8.2580\{14\}~\cite{ARCv3_2020} & $-$3.9175\{14\}~\cite{ARCv3_2020} & $-$4.6961\{14\}~\cite{ARCv3_2020} & 3.2027\{15\}~\cite{ARCv3_2020} & 2.5319\{15\}~\cite{ARCv3_2020}  \\
\hline\hline
\end{tabular}
\end{scriptsize}
\end{table*}

\begin{table*}[tp] 
\caption{\label{tab_polar_tensor} Selected tensor polarizabilities ($\alpha_2$) of the $nP_{3/2}$, $nD_J$, and $nF_J$ states of Cs in a.u. The notation $\{x\}$ denotes $\times 10^x$.}
\centering
\begin{scriptsize}
\begin{tabular}{clllll}
\hline\hline            
$n$ & \multicolumn{1}{c}{$\alpha_2(P_{3/2})$} 
    & \multicolumn{1}{c}{$\alpha_2(D_{3/2})$}
    & \multicolumn{1}{c}{$\alpha_2(D_{5/2})$}   
    & \multicolumn{1}{c}{$\alpha_2(F_{5/2})$}
    & \multicolumn{1}{c}{$\alpha_2(F_{7/2})$}\\ \hline
8 & $-$3.0935\{4\} & 3.4037\{5\} & 6.8207\{5\} & $-$1.9336\{7\} & $-$2.2403\{7\}  \\
 & $-$3.3450\{4\}~\cite{ARCv3_2020} & 3.7136\{5\}~\cite{ARCv3_2020} & 7.4357\{5\}~\cite{ARCv3_2020} & $-$2.1208\{7\}~\cite{ARCv3_2020} & $-$2.4505\{7\}~\cite{ARCv3_2020}  \\
 & $-$3.057(41)\{4\}~\cite{Safronova2016} & 3.386(31)\{5\}~\cite{Safronova2016} & 6.781(49)\{5\}~\cite{Safronova2016} &  &   \\
 & $-$3.02\{4\}~\cite{Wijngaarden_1994} & 3.36\{5\}~\cite{Wijngaarden_1994} & 6.75\{5\}~\cite{Wijngaarden_1994} & \dls 7.87\{5\}~\cite{Wijngaarden_1994} & \dls 1.03\{6\}~\cite{Wijngaarden_1994}  \\ \hline
10 & $-$4.5170\{5\} & 3.4204\{6\} & 6.8758\{6\} & $-$1.0029\{8\} & $-$1.1580\{8\}  \\
 & $-$4.9337\{5\}~\cite{ARCv3_2020} & 3.7455\{6\}~\cite{ARCv3_2020} & 7.5322\{6\}~\cite{ARCv3_2020} & $-$1.0969\{8\}~\cite{ARCv3_2020} & $-$9.8166\{7\}~\cite{ARCv3_2020}  \\
 & $-$4.511(48)\{5\}~\cite{Safronova2016} & 3.419(22)\{6\}~\cite{Safronova2016}  & 6.871(35)\{6\}~\cite{Safronova2016} &  &   \\
 & $-$4.49\{5\}~\cite{Wijngaarden_1994}  & 3.41\{6\}~\cite{Wijngaarden_1994} & 6.85\{6\}~\cite{Wijngaarden_1994} & \dls 3.73\{6\}~\cite{Wijngaarden_1994} & \dls 4.94\{6\}~\cite{Wijngaarden_1994}  \\
 & $-$4.60\{5\}~\cite{Yerokhin_2016} & 3.47\{6\}~\cite{Yerokhin_2016} & 7.01\{6\}~\cite{Yerokhin_2016} &  &   \\ \hline
12 & $-$3.0439\{6\} & 1.9016\{7\} & 3.8244\{7\} & $-$3.7348\{8\} & $-$4.3120\{8\}  \\
 & $-$3.3300\{6\}~\cite{ARCv3_2020} & 2.0860\{7\}~\cite{ARCv3_2020} & 4.1948\{7\}~\cite{ARCv3_2020} & $-$4.0875\{8\}~\cite{ARCv3_2020} & $-$3.6474\{8\}~\cite{ARCv3_2020}  \\
 & $-$3.041(24)\{6\}~\cite{Safronova2016} & 1.8734(158)\{7\}~\cite{Safronova2016} & 3.8424(291)\{7\}~\cite{Safronova2016} &  &   \\
 & $-$3.05\{6\}~\cite{Wijngaarden_1994} & 1.91\{7\}~\cite{Wijngaarden_1994} & 3.83\{7\}~\cite{Wijngaarden_1994} &  &   \\
 & $-$3.08\{6\}~\cite{Yerokhin_2016} & 1.92\{7\}~\cite{Yerokhin_2016} & 3.87\{7\}~\cite{Yerokhin_2016} &  &   \\ \hline
14 & $-$1.3476\{7\} & 7.4856\{7\} & 1.5049\{8\} & $-$1.1230\{9\} & $-$1.2965\{9\}  \\
 & $-$1.4775\{7\}~\cite{ARCv3_2020} & 8.2071\{7\}~\cite{ARCv3_2020} & 1.6503\{8\}~\cite{ARCv3_2020} & $-$1.2294\{9\}~\cite{ARCv3_2020} & $-$1.0951\{9\}~\cite{ARCv3_2020}  \\
 & $-$1.36\{7\}~\cite{Yerokhin_2016} & 7.53\{7\}~\cite{Yerokhin_2016} & 1.52\{8\}~\cite{Yerokhin_2016} &  &   \\
 & $-$1.36\{7\}~\cite{Wijngaarden_1994} &  &  &  &   \\ \hline
16 & $-$4.5724\{7\} & 2.3464\{8\} & 4.7139\{8\} & $-$2.8992\{9\} & $-$3.3467\{9\}  \\
 & $-$5.0135\{7\}~\cite{ARCv3_2020} & 2.5716\{8\}~\cite{ARCv3_2020} & 5.1693\{8\}~\cite{ARCv3_2020} & $-$3.1743\{9\}~\cite{ARCv3_2020} & $-$2.8242\{9\}~\cite{ARCv3_2020}  \\
 & $-$4.60\{7\}~\cite{Yerokhin_2016} & 2.35\{8\}~\cite{Yerokhin_2016} & 4.74\{8\}~\cite{Yerokhin_2016} &  &   \\ \hline
20 & $-$3.1801\{8\} & 1.4762\{9\} & 2.9617\{9\} & $-$1.4043\{10\} & $-$1.6208\{10\}  \\
 & $-$3.4875\{8\}~\cite{ARCv3_2020} & 1.6173\{9\}~\cite{ARCv3_2020} & 3.2479\{9\}~\cite{ARCv3_2020} & $-$1.5376\{10\}~\cite{ARCv3_2020} & $-$1.3660\{10\}~\cite{ARCv3_2020}  \\ \hline
25 & $-$2.0141\{9\} & 8.6957\{9\} & 1.7422\{10\} & $-$6.7626\{10\} & $-$7.8040\{10\}  \\
 & $-$2.2092\{9\}~\cite{ARCv3_2020} & 9.5247\{9\}~\cite{ARCv3_2020} & 1.9107\{10\}~\cite{ARCv3_2020} & $-$7.4043\{10\}~\cite{ARCv3_2020} & $-$6.5706\{10\}~\cite{ARCv3_2020}  \\
 &  & 8.76(36)\{9\}~\cite{Lei_1995} & 1.87(5)\{10\}~\cite{Lei_1995} &  &   \\ \hline
30 & $-$8.6510\{9\} & 3.5706\{10\} & 7.1461\{10\} & $-$2.4360\{11\} & $-$2.8109\{11\}  \\
 & $-$9.4902\{9\}~\cite{ARCv3_2020} & 3.9106\{10\}~\cite{ARCv3_2020} & 7.8377\{10\}~\cite{ARCv3_2020} & $-$2.6670\{11\}~\cite{ARCv3_2020} & $-$2.3652\{11\}~\cite{ARCv3_2020}  \\
 &  & 3.61(8)\{10\}~\cite{Lei_1995} & 6.99(12)\{10\}~\cite{Lei_1995} &  &   \\ \hline
35 & $-$2.8863\{10\} & 1.1553\{11\} & 2.3103\{11\} & $-$7.1892\{11\} & $-$8.2952\{11\}  \\
 & $-$3.1666\{10\}~\cite{ARCv3_2020} & 1.2653\{11\}~\cite{ARCv3_2020} & 2.5340\{11\}~\cite{ARCv3_2020} & $-$7.8709\{11\}~\cite{ARCv3_2020} & $-$6.9768\{11\}~\cite{ARCv3_2020}  \\
 &  & 1.25(5)\{11\}~\cite{Lei_1995} & 2.35(4)\{11\}~\cite{Lei_1995} &  &   \\ \hline
37 & $-$4.4334\{10\} & 1.7575\{11\} & 3.5135\{11\} & $-$1.0617\{12\} & $-$1.2251\{12\}  \\
 & $-$4.8639\{10\}~\cite{ARCv3_2020} & 1.9246\{11\}~\cite{ARCv3_2020} & 3.8537\{11\}~\cite{ARCv3_2020} & $-$1.1624\{12\}~\cite{ARCv3_2020} & $-$1.0302\{12\}~\cite{ARCv3_2020} \\
 & $-$4.43\{10\}~\cite{Yerokhin_2016} & 1.76\{11\}~\cite{Yerokhin_2016} & 3.51\{11\}~\cite{Yerokhin_2016} & $-$1.06\{12\}~\cite{Bai_2024} & $-$1.23\{12\}~\cite{Bai_2024} \\ \hline
39 & $-$6.6424\{10\} & 2.6107\{11\} & 5.2178\{11\} & $-$1.5360\{12\} & $-$1.7723\{12\}  \\
 & $-$7.2877\{10\}~\cite{ARCv3_2020} & 2.8589\{11\}~\cite{ARCv3_2020} & 5.7231\{11\}~\cite{ARCv3_2020} & $-$1.6816\{12\}~\cite{ARCv3_2020} & $-$1.4902\{12\}~\cite{ARCv3_2020}  \\
 & $-$6.64\{10\}~\cite{Yerokhin_2016} & 2.61\{11\}~\cite{Yerokhin_2016} & 5.21\{11\}~\cite{Yerokhin_2016} &  &   \\
 &  & 3.0(2)\{11\}~\cite{Zhao_2011} & 5.59(12)\{11\}~\cite{Zhao_2011} &  &   \\
 &  &  & 5.22\{11\}~\cite{Bai_2024}  &  &   \\ \hline
40 & $-$8.0622\{10\} & 3.1562\{11\} & 6.3074\{11\} & $-$1.8345\{12\} & $-$2.1166\{12\}  \\
 & $-$8.8455\{10\}~\cite{ARCv3_2020} & 3.4563\{11\}~\cite{ARCv3_2020} & 6.9182\{11\}~\cite{ARCv3_2020} & $-$2.0084\{12\}~\cite{ARCv3_2020} & $-$1.7797\{12\}~\cite{ARCv3_2020}  \\
 &  & 3.33(7)\{11\}~\cite{Lei_1995} & 6.55(20)\{11\}~\cite{Lei_1995} &  &   \\ \hline
42 & $-$1.1695\{11\} & 4.5455\{11\} & 9.0816\{11\} & $-$2.5829\{12\} & $-$2.9802\{12\}  \\
 & $-$1.2832\{11\}~\cite{ARCv3_2020} & 4.9776\{11\}~\cite{ARCv3_2020} & 9.9611\{11\}~\cite{ARCv3_2020} & $-$2.8278\{12\}~\cite{ARCv3_2020} & $-$2.5055\{12\}~\cite{ARCv3_2020}  \\
 & $-$1.17\{11\}~\cite{Yerokhin_2016} & 4.54\{11\}~\cite{Yerokhin_2016} & 9.07\{11\}~\cite{Yerokhin_2016} & &   \\
 &  & 4.66(40)\{11\}~\cite{Zhao_2011} & 9.24(84)\{11\}~\cite{Zhao_2011} &  &   \\ \hline
45 & $-$1.9738\{11\} & 7.5979\{11\} & 1.5176\{12\} & $-$4.1898\{12\} & $-$4.8341\{12\}  \\
 & $-$2.1656\{11\}~\cite{ARCv3_2020} & 8.3201\{11\}~\cite{ARCv3_2020} & 1.6646\{12\}~\cite{ARCv3_2020} & $-$4.5869\{12\}~\cite{ARCv3_2020} & $-$4.0635\{12\}~\cite{ARCv3_2020}  \\
 &  & 9.16(36)\{11\}~\cite{Lei_1995} & 1.72(4)\{12\}~\cite{Lei_1995} &  &   \\ \hline
50 & $-$4.3645\{11\} & 1.6580\{12\} & 3.3102\{12\} & $-$8.7687\{12\} & $-$1.0117\{13\}  \\
 & $-$4.7888\{11\}~\cite{ARCv3_2020} & 1.8156\{12\}~\cite{ARCv3_2020} & 3.6309\{12\}~\cite{ARCv3_2020} & $-$9.5996\{12\}~\cite{ARCv3_2020} & $-$8.5027\{12\}~\cite{ARCv3_2020}  \\
 & $-$4.36\{11\}~\cite{Yerokhin_2016,Bhowmik_2014} & 1.66\{12\}~\cite{Yerokhin_2016,Bhowmik_2014} & 3.31\{12\}~\cite{Yerokhin_2016,Bhowmik_2014}  & &  \\
 &  & 2.07(6)\{12\}~\cite{Lei_1995} & 3.77(8)\{12\}~\cite{Lei_1995} &  &   \\
 &  & 1.67(7)\{12\}~\cite{Zhao_2011} & 2.67(23)\{12\}~\cite{Zhao_2011} &  &   \\ \hline
55 & $-$8.9004\{11\} & 3.3453\{12\} & 6.6765\{12\} & $-$1.7100\{13\} & $-$1.9729\{13\}  \\
 & $-$9.7659\{11\}~\cite{ARCv3_2020} & 3.6632\{12\}~\cite{ARCv3_2020} & 7.3233\{12\}~\cite{ARCv3_2020} & $-$1.8721\{13\}~\cite{ARCv3_2020} & $-$1.6579\{13\}~\cite{ARCv3_2020}  \\
 &  & 4.86(20)\{12\}~\cite{Lei_1995} & 8.96(20)\{12\}~\cite{Lei_1995} &  &   \\ \hline
60 & $-$1.6992\{12\} & 6.3309\{12\} & 1.2631\{13\} & $-$3.1461\{13\} & $-$3.6297\{13\}  \\
 & $-$1.8644\{12\}~\cite{ARCv3_2020} & 6.9323\{12\}~\cite{ARCv3_2020} & 1.3855\{13\}~\cite{ARCv3_2020} & $-$3.4442\{13\}~\cite{ARCv3_2020} & $-$3.0497\{13\}~\cite{ARCv3_2020}  \\ 
 & $-$1.90\{12\}~\cite{Bhowmik_2014} & 6.17\{12\}~\cite{Bhowmik_2014} & 1.30\{13\}~\cite{Bhowmik_2014} & &  \\ \hline
65 & $-$3.0711\{12\} & 1.1359\{13\} & 2.2657\{13\} & $-$5.5119\{13\} & $-$6.3592\{13\}  \\
 & $-$3.3698\{12\}~\cite{ARCv3_2020} & 1.2438\{13\}~\cite{ARCv3_2020} & 2.4852\{13\}~\cite{ARCv3_2020} & $-$6.0341\{13\}~\cite{ARCv3_2020} & $-$5.3425\{13\}~\cite{ARCv3_2020}  \\ \hline
70 & $-$5.3000\{12\} & 1.9481\{13\} & 3.8849\{13\} & $-$9.2629\{13\} & $-$1.0687\{14\}  \\
 & $-$5.8156\{12\}~\cite{ARCv3_2020} & 2.1331\{13\}~\cite{ARCv3_2020} & 4.2614\{13\}~\cite{ARCv3_2020} & $-$1.0140\{14\}~\cite{ARCv3_2020} & $-$8.9773\{13\}~\cite{ARCv3_2020}  \\
 & $-$5.05\{12\}~\cite{Bhowmik_2014} & 2.10\{13\}~\cite{Bhowmik_2014} & 3.94\{13\}~\cite{Bhowmik_2014} & &  \\ \hline
75 & $-$8.7924\{12\} & 3.2145\{13\} & 6.4092\{13\} & $-$1.5018\{14\} & $-$1.7326\{14\}  \\
 & $-$9.6478\{12\}~\cite{ARCv3_2020} & 3.5198\{13\}~\cite{ARCv3_2020} & 7.0302\{13\}~\cite{ARCv3_2020} & $-$1.6441\{14\}~\cite{ARCv3_2020} & $-$1.4554\{14\}~\cite{ARCv3_2020}  \\
 & $-$8.8068\{12\}~\cite{Hou_2025} &  &  &  &   \\ \hline
80 & $-$1.4096\{13\} & 5.1297\{13\} & 1.0226\{14\} & $-$2.3600\{14\} & $-$2.7227\{14\}  \\
 & $-$1.5468\{13\}~\cite{ARCv3_2020} & 5.6168\{13\}~\cite{ARCv3_2020} & 1.1217\{14\}~\cite{ARCv3_2020} & $-$2.5836\{14\}~\cite{ARCv3_2020} & $-$2.2869\{14\}~\cite{ARCv3_2020}  \\
 & $-$1.4145\{13\}~\cite{Hou_2025} &  &  &  &   \\ \hline
85 & $-$2.1935\{13\} & 7.9497\{13\} & 1.5845\{14\} & $-$3.6083\{14\} & $-$4.1629\{14\}  \\
 & $-$2.4069\{13\}~\cite{ARCv3_2020} & 8.7046\{13\}~\cite{ARCv3_2020} & 1.7380\{14\}~\cite{ARCv3_2020} & $-$3.9501\{14\}~\cite{ARCv3_2020} & $-$3.4963\{14\}~\cite{ARCv3_2020}  \\
 & $-$2.2279\{13\}~\cite{Hou_2025} &  &  &  &   \\ \hline
90 & $-$3.3247\{13\} & 1.2006\{14\} & 2.3927\{14\} & $-$5.3845\{14\} & $-$6.2119\{14\}  \\
 & $-$3.6482\{13\}~\cite{ARCv3_2020} & 1.3146\{14\}~\cite{ARCv3_2020} & 2.6245\{14\}~\cite{ARCv3_2020} & $-$5.8944\{14\}~\cite{ARCv3_2020} & $-$5.2170\{14\}~\cite{ARCv3_2020}  \\ \hline
95 & $-$4.9230\{13\} & 1.7721\{14\} & 3.5311\{14\} & $-$7.8627\{14\} & $-$9.0710\{14\}  \\
 & $-$5.4021\{13\}~\cite{ARCv3_2020} & 1.9403\{14\}~\cite{ARCv3_2020} & 3.8733\{14\}~\cite{ARCv3_2020} & $-$8.6073\{14\}~\cite{ARCv3_2020} & $-$7.6178\{14\}~\cite{ARCv3_2020}  \\ \hline
100 & $-$7.1392\{13\} & 2.5624\{14\} & 5.1055\{14\} & $-$1.1260\{15\} & $-$1.2991\{15\}  \\
 & $-$7.8340\{13\}~\cite{ARCv3_2020} & 2.8057\{14\}~\cite{ARCv3_2020} & 5.6002\{14\}~\cite{ARCv3_2020} & $-$1.2327\{15\}~\cite{ARCv3_2020} & $-$1.0909\{15\}~\cite{ARCv3_2020}  \\
\hline\hline
\end{tabular}
\end{scriptsize}
\end{table*}

The polarizabilities of Rydberg states scale as $n^7$~\cite{Gallagher_1994,Kamenski_2014}. 
They can be expanded as a~series in powers of $[n-\delta^{(l,J)}]$ as follows~\cite{Kamenski_2014, Pawlak_2024}  
\begin{equation}\label{eqn:coeff}
\alpha_{i}^{(n,l,J)} =  \left [ n-\delta^{(l,J)}(n) \right]^{\!7}  \left(\eta^{(l,J)}_{i,0} + \sum_{p=1}^{\infty} \frac{\eta_{i,p}^{(l,J)}}{\left [ n-\delta^{(l,J)}(n) \right ]^p}  \right)\!,     
\end{equation}
where $i=0,2$. We determined the $\eta^{(l,J)}_{i,p}$ coefficients for all series under consideration by fitting our calculated polarizabilities to Eq.~(\ref{eqn:coeff}) up to $p=2$. They are presented in Table~\ref{tab:coeff}. One can readily reproduce $\alpha_0$ and $\alpha_2$ with four or five significant figures accuracy for high-lying levels.

\section{Conclusions}

We presented high-precision, absolute frequency measurements of Cs energy levels from the $\vert 6S_{1/2}, F=3 \rangle$ hyperfine ground state to $nF_{5/2}(n=28$--$68)$ and $nF_{7/2}(n=28$--$68)$ Rydberg states with an accuracy of $< 60\,\rm kHz$. We determined the quantum defects of the Cs $nF_{5/2}$ and $nF_{7/2}$ series as well as the ionization energy for each series by fitting the absolute frequency measurements to the modified Ritz formula. The extracted ionization energy is $31406.467 751 52(25)$~cm$^{-1}$ for the $nF_{5/2}$ series and $31406.467 751 46(26)$~cm$^{-1}$ for the $nF_{7/2}$ series. Both ionization energies agree with our recent measurements, $31 406.467 751 48 (14)$~cm$^{-1}$, based on the $nS_{1/2}$ and $nD_J$ series~\cite{Shen2024}. We compared our quantum defects with previously reported quantum defects for the $nF_J$ series and found that the quantum defect expansion can be truncated after the third term at the current level of precision. We estimated the individual contributions of core polarization and core penetration to the quantum defects and compared their relative significance. For the $nS_{1/2}$ series, the contribution from core penetration is higher than the one from core polarization by a~factor of 30, while for the $nF_J$ series, the contribution from core polarization is two orders of magnitude larger than the contribution from core penetration. The dominance of the core polarization for the $nF_J$ series demonstrates the dramatic reduction in the core penetration effect with increasing $l$. We also updated the fine structure coefficients of the $nF_J$ series and improved the uncertainty by a factor of 2 from previous work. Using new Cs Rydberg wave functions computed numerically by the Numerov method with the quantum defect energies, we calculated the reduced electric-dipole matrix elements for the $nD_J \leftrightarrow n^{\prime} F_{J^\prime}$ transitions. The matrix elements are found to be in accord with high level, many-body relativistic calculations at low~$n$, where our measurements were extrapolated. Finally, we calculated the scalar and tensor polarizabilities of the selected $nS_{1/2}$, $nP_J$, $nD_J$, and $nF_J$ states and contrasted them with prior experimental and theoretical works. We represented the polarizability as a series in powers of the effective principal quantum number and determined the main coefficients of the expansion. One can easily evaluate $\alpha_0$ and $\alpha_2$ for quantum numbers $n$, $l<4$, $J$. The precision measurements of the Cs energy levels and the accurately calculated scalar and tensor polarizabilites presented in this work will be beneficial not only for fundamental physics but also for practical applications of atom-based quantum technologies, such as Rydberg atom-based electric field sensors and neutral atom-based quantum computing.
\\

\section*{Acknowledgments}
This work has been supported by The National Research Council Internet of Things: Quantum Sensors Challenge program through Contract No. QSP-058-1. The calculations were supported at ITAMP by a grant from the U.S. National Science Foundation.

\end{document}